\input harvmac.tex
 \input epsf.tex
 \input amssym



\def\figin{\epsfcheck\figin}\def\figins{\epsfcheck\figins}
\def\epsfcheck{\ifx\epsfbox\UnDeFiNeD
\message{(NO epsf.tex, FIGURES WILL BE IGNORED)}
\gdef\figin##1{\vskip2in}\gdef\figins##1{\hskip.5in}
\else\message{(FIGURES WILL BE INCLUDED)}%
\gdef\figin##1{##1}\gdef\figins##1{##1}\fi}
\def\DefWarn#1{}
\def\figinsert{\goodbreak\midinsert}
\def\ifig#1#2#3{\DefWarn#1\xdef#1{fig.~\the\figno}
\writedef{#1\leftbracket fig.\noexpand~\the\figno} %
\figinsert\figin{\centerline{#3}}\medskip\centerline{\vbox{\baselineskip12pt
\advance\hsize by -1truein\noindent\footnotefont{\bf
Fig.~\the\figno:} #2}}
\bigskip\endinsert\global\advance\figno by1}



\def \eqna  {\begin{eqnarray}}
\def \eeq  {\end{eqnarray}}


\def\frac#1#2{{#1 \over #2}}
\def\text#1{#1}


\lref\RyuBV{
  S.~Ryu and T.~Takayanagi,
Phys.\ Rev.\ Lett.\  {\bf 96}, 181602 (2006).
[hep-th/0603001].
}

\lref\HubenyXT{
  V.~E.~Hubeny, M.~Rangamani and T.~Takayanagi,
JHEP {\bf 0707}, 062 (2007).
[arXiv:0705.0016 [hep-th]].
}

\lref\AbajoArrastiaYT{
  J.~Abajo-Arrastia, J.~Aparicio and E.~Lopez,
  ``Holographic Evolution of Entanglement Entropy,''
JHEP {\bf 1011}, 149 (2010).
[arXiv:1006.4090 [hep-th]].
}

\lref\AparicioZY{
  J.~Aparicio and E.~Lopez,
  ``Evolution of Two-Point Functions from Holography,''
JHEP {\bf 1112}, 082 (2011).
[arXiv:1109.3571 [hep-th]].
}

\lref\OstlundZZ{
  S.~Ostlund and S.~Rommer,
  ``Thermodynamic Limit of Density Matrix Renormalization for the spin-1 heisenberg chain,''
Phys.\ Rev.\ Lett.\  {\bf 75}, 3537 (1995).
[cond-mat/9503107].
}

\lref\cirac{
F. Verstraete and J. I. Cirac,
``Renormalization algorithms for Quantum-Many Body Systems in two and higher dimensions,"
[arXiv:cond36
mat/0407066v1] (2004).
}

\lref\AlbashMV{
  T.~Albash and C.~V.~Johnson,
  ``Evolution of Holographic Entanglement Entropy after Thermal and Electromagnetic Quenches,''
New J.\ Phys.\  {\bf 13}, 045017 (2011).
[arXiv:1008.3027 [hep-th]].
}

\lref\TakayanagiWP{
  T.~Takayanagi and T.~Ugajin,
  ``Measuring Black Hole Formations by Entanglement Entropy via Coarse-Graining,''
JHEP {\bf 1011}, 054 (2010).
[arXiv:1008.3439 [hep-th]].
}

\lref\MaldacenaKR{
  J.~M.~Maldacena,
JHEP {\bf 0304}, 021 (2003).
[hep-th/0106112].
}
\lref\BakJM{
  D.~Bak, M.~Gutperle and S.~Hirano,
JHEP {\bf 0702}, 068 (2007).
[hep-th/0701108].
}
\lref\BakQW{
  D.~Bak, M.~Gutperle and A.~Karch,
JHEP {\bf 0712}, 034 (2007).
[arXiv:0708.3691 [hep-th]].
}

\lref\CalabreseEU{
  P.~Calabrese and J.~L.~Cardy,
J.\ Stat.\ Mech.\  {\bf 0406}, P06002 (2004).
[hep-th/0405152].
}
\lref\IsraelUR{
  W.~Israel,
Phys.\ Lett.\ A {\bf 57}, 107 (1976)..
}

\lref\NishiokaUN{
  T.~Nishioka, S.~Ryu and T.~Takayanagi,
J.\ Phys.\ A {\bf 42}, 504008 (2009).
[arXiv:0905.0932 [hep-th]].
}

\lref\BelinDVA{
  A.~Belin, A.~Maloney and S.~Matsuura,
[arXiv:1306.2640 [hep-th]].
}

\lref\CalabreseIN{
  P.~Calabrese and J.~L.~Cardy,
J.\ Stat.\ Mech.\  {\bf 0504}, P04010 (2005).
[cond-mat/0503393].
}
\lref\CalabreseRG{
  P.~Calabrese and J.~Cardy,
J.\ Stat.\ Mech.\  {\bf 0706}, P06008 (2007).
[arXiv:0704.1880 [cond-mat.stat-mech]].
}

\lref\FursaevIX{
  D.~V.~Fursaev,
Phys.\ Rev.\ D {\bf 82}, 064013 (2010), [Erratum-ibid.\ D {\bf 86}, 049903 (2012)].
[arXiv:1006.2623 [hep-th]].
}

\lref\CalabreseRX{
  P.~Calabrese and J.~L.~Cardy,
  ``Time-dependence of correlation functions following a quantum quench,''
Phys.\ Rev.\ Lett.\  {\bf 96}, 136801 (2006).
[cond-mat/0601225].
}

\lref\CalabreseQY{
  P.~Calabrese and J.~Cardy,
  ``Entanglement entropy and conformal field theory,''
J.\ Phys.\ A {\bf 42}, 504005 (2009).
[arXiv:0905.4013 [cond-mat.stat-mech]].
}

\lref\MorrisonIZ{
  I.~A.~Morrison and M.~M.~Roberts,
  ``Mutual information between thermo-field doubles and disconnected holographic boundaries,''
[arXiv:1211.2887 [hep-th]].
}

\lref\BanadosWN{
  M.~Banados, C.~Teitelboim and J.~Zanelli,
 ``The Black hole in three-dimensional space-time,''
Phys.\ Rev.\ Lett.\  {\bf 69}, 1849 (1992).
[hep-th/9204099].
}

\lref\MaldacenaBW{
  J.~M.~Maldacena and A.~Strominger,
  ``AdS(3) black holes and a stringy exclusion principle,''
JHEP {\bf 9812}, 005 (1998).
[hep-th/9804085].
}

\lref\HeadrickZT{
  M.~Headrick,
Phys.\ Rev.\ D {\bf 82}, 126010 (2010).
[arXiv:1006.0047 [hep-th]].
}

\lref\SiopsisUP{
  G.~Siopsis,
  ``Large mass expansion of quasinormal modes in AdS(5),''
Phys.\ Lett.\ B {\bf 590}, 105 (2004).
[hep-th/0402083].
}

\lref\LoukoHC{
  J.~Louko and D.~Marolf,
  ``Single exterior black holes and the AdS / CFT conjecture,''
Phys.\ Rev.\ D {\bf 59}, 066002 (1999).
[hep-th/9808081].
}

\lref\HartmanQMA{
  T.~Hartman and J.~Maldacena,
[arXiv:1303.1080 [hep-th]].
}
\lref\FreivogelQH{
  B.~Freivogel, V.~E.~Hubeny, A.~Maloney, R.~C.~Myers, M.~Rangamani and S.~Shenker,
JHEP {\bf 0603}, 007 (2006).
[hep-th/0510046].
}
\lref\CarlipSA{
  S.~Carlip and C.~Teitelboim,
Class.\ Quant.\ Grav.\  {\bf 12}, 1699 (1995).
[gr-qc/9312002].
}

\lref\FestucciaZX{
  G.~Festuccia and H.~Liu,
  ``A Bohr-Sommerfeld quantization formula for quasinormal frequencies of AdS black holes,''
Adv.\ Sci.\ Lett.\  {\bf 2}, 221 (2009).
[arXiv:0811.1033 [gr-qc]].
}

\lref\MaldacenaXP{
  J.~Maldacena and G.~L.~Pimentel,
  ``Entanglement entropy in de Sitter space,''
JHEP {\bf 1302}, 038 (2013).
[arXiv:1210.7244 [hep-th]].
}

\lref\BriganteJV{
  M.~Brigante, G.~Festuccia and H.~Liu,
  ``Hagedorn divergences and tachyon potential,''
JHEP {\bf 0706}, 008 (2007).
[hep-th/0701205].
}

\lref\FestucciaSA{
  G.~Festuccia and H.~Liu,
  ``The Arrow of time, black holes, and quantum mixing of large N Yang-Mills theories,''
JHEP {\bf 0712}, 027 (2007).
[hep-th/0611098].
}

\lref\BriganteBQ{
  M.~Brigante, G.~Festuccia and H.~Liu,
  ``Inheritance principle and non-renormalization theorems at finite temperature,''
Phys.\ Lett.\ B {\bf 638}, 538 (2006).
[hep-th/0509117].
}

\lref\FestucciaPI{
  G.~Festuccia and H.~Liu,
  ``Excursions beyond the horizon: Black hole singularities in Yang-Mills theories. I.,''
JHEP {\bf 0604}, 044 (2006).
[hep-th/0506202].
}

\lref\FidkowskiNF{
  L.~Fidkowski, V.~Hubeny, M.~Kleban and S.~Shenker,
  ``The Black hole singularity in AdS / CFT,''
JHEP {\bf 0402}, 014 (2004).
[hep-th/0306170].
}

\lref\SwingleBG{
  B.~Swingle,
  ``Entanglement Renormalization and Holography,''
Phys.\ Rev.\ D {\bf 86}, 065007 (2012).
[arXiv:0905.1317 [cond-mat.str-el]].
}

\lref\BalasubramanianCE{
  V.~Balasubramanian, A.~Bernamonti, J.~de Boer, N.~Copland, B.~Craps, E.~Keski-Vakkuri, B.~Muller and A.~Schafer {\it et al.},
  ``Thermalization of Strongly Coupled Field Theories,''
Phys.\ Rev.\ Lett.\  {\bf 106}, 191601 (2011).
[arXiv:1012.4753 [hep-th]].
}

\lref\BalasubramanianUR{
  V.~Balasubramanian, A.~Bernamonti, J.~de Boer, N.~Copland, B.~Craps, E.~Keski-Vakkuri, B.~Muller and A.~Schafer {\it et al.},
  ``Holographic Thermalization,''
Phys.\ Rev.\ D {\bf 84}, 026010 (2011).
[arXiv:1103.2683 [hep-th]].
}

\lref\AsplundCQ{
  C.~T.~Asplund and S.~G.~Avery,
  ``Evolution of Entanglement Entropy in the D1-D5 Brane System,''
Phys.\ Rev.\ D {\bf 84}, 124053 (2011).
[arXiv:1108.2510 [hep-th]].
}

\lref\BasuFT{
  P.~Basu and S.~R.~Das,
  ``Quantum Quench across a Holographic Critical Point,''
JHEP {\bf 1201}, 103 (2012).
[arXiv:1109.3909 [hep-th]].
}

\lref\BasuGG{
  P.~Basu, D.~Das, S.~R.~Das and T.~Nishioka,
  ``Quantum Quench Across a Zero Temperature Holographic Superfluid Transition,''
[arXiv:1211.7076 [hep-th]].
}

\lref\BalasubramanianAT{
  V.~Balasubramanian, A.~Bernamonti, N.~Copland, B.~Craps and F.~Galli,
  ``Thermalization of mutual and tripartite information in strongly coupled two dimensional conformal field theories,''
Phys.\ Rev.\ D {\bf 84}, 105017 (2011).
[arXiv:1110.0488 [hep-th]].
}

\lref\AllaisYS{
  A.~Allais and E.~Tonni,
  ``Holographic evolution of the mutual information,''
JHEP {\bf 1201}, 102 (2012).
[arXiv:1110.1607 [hep-th]].
}

\lref\HubenyRY{
  V.~E.~Hubeny,
  ``Extremal surfaces as bulk probes in AdS/CFT,''
JHEP {\bf 1207}, 093 (2012).
[arXiv:1203.1044 [hep-th]].
}

\lref\MaldacenaXP{
  J.~Maldacena and G.~L.~Pimentel,
  ``Entanglement entropy in de Sitter space,''
JHEP {\bf 1302}, 038 (2013).
[arXiv:1210.7244 [hep-th]].
}

\lref\BuchelLLA{
  A.~Buchel, L.~Lehner, R.~C.~Myers and A.~van Niekerk,
  ``Quantum quenches of holographic plasmas,''
[arXiv:1302.2924 [hep-th]].
}
\lref\VidalMera{
G.~Vidal,
``A class of quantum many-body states that can be efficiently simulated,"
Phys.~Rev.~Lett.~101, 110501 (2008).
[arXiv: quant-ph/0610099].
}

\lref\TakayanagiZK{
  T.~Takayanagi,
  ``Holographic Dual of BCFT,''
Phys.\ Rev.\ Lett.\  {\bf 107}, 101602 (2011).
[arXiv:1105.5165 [hep-th]].
}

\lref\BoussoSJ{
  R.~Bousso, S.~Leichenauer and V.~Rosenhaus,
  ``Light-sheets and AdS/CFT,''
Phys.\ Rev.\ D {\bf 86}, 046009 (2012).
[arXiv:1203.6619 [hep-th]].
}

\lref\CzechBH{
  B.~Czech, J.~L.~Karczmarek, F.~Nogueira and M.~Van Raamsdonk,
  ``The Gravity Dual of a Density Matrix,''
Class.\ Quant.\ Grav.\  {\bf 29}, 155009 (2012).
[arXiv:1204.1330 [hep-th]].
}

\lref\BoussoMH{
  R.~Bousso, B.~Freivogel, S.~Leichenauer, V.~Rosenhaus and C.~Zukowski,
  ``Null Geodesics, Local CFT Operators and AdS/CFT for Subregions,''
[arXiv:1209.4641 [hep-th]].
}

\lref\VidalHolo{
 G.~Evenbly and G.~Vidal,
 ``Tensor network states and geometry,"
J.~Stat.~Phys.~{\bf 145}, 891 (2011).
[arXiv: 1106.1082 [quant-ph]].
}

\lref\VidalRev{
G.~Vidal,
``Entanglement Renormalization: an introduction,"
in {\it Understanding Quantum Phase Transitions}, L.~D.~Carr ed. (Taylor \& Francis, Boca Raton, 2010).
[arXiv: 0912.1651 [cond-mat.str-el]].
}

\lref\SchollRev{
U.~Schollwoeck,
``The density-matrix renormalization group in the age of matrix product states,"
Ann.~Phys.~{\bf 326}, 96 (2011).
[arXiv:1008.3477 [cond-mat.str-el]].
}

\lref\VidalTree{
Y.~Shi, L.~Duan, G.~Vidal,
``Classical simulation of quantum many-body systems with a tree tensor network,"
Phys.~Rev.~A {\bf 74}, 022320 (2006).
[arXiv: quant-ph/0511070].
}

\lref\UnruhHY{
  W.~G.~Unruh, G.~Hayward, W.~Israel and D.~Mcmanus,
Phys.\ Rev.\ Lett.\  {\bf 62}, 2897 (1989)..
}

\lref\SwingleWQ{
  B.~Swingle,
  ``Constructing holographic spacetimes using entanglement renormalization,''
[arXiv:1209.3304 [hep-th]].
}

\lref\NozakiZJ{
  M.~Nozaki, S.~Ryu and T.~Takayanagi,
  ``Holographic Geometry of Entanglement Renormalization in Quantum Field Theories,''
[arXiv:1208.3469 [hep-th]].
}

\lref\WhiteZZ{
  S.~R.~White,
  ``Density matrix formulation for quantum renormalization groups,''
Phys.\ Rev.\ Lett.\  {\bf 69}, 2863 (1992)..
}

\lref\HolzheyWE{
  C.~Holzhey, F.~Larsen and F.~Wilczek,
  ``Geometric and renormalized entropy in conformal field theory,''
Nucl.\ Phys.\ B {\bf 424}, 443 (1994).
[hep-th/9403108].
}

\lref\CalabreseRX{
  P.~Calabrese and J.~L.~Cardy,
  ``Time-dependence of correlation functions following a quantum quench,''
Phys.\ Rev.\ Lett.\  {\bf 96}, 136801 (2006).
[cond-mat/0601225].
}

\lref\CallanPY{
  C.~G.~Callan, Jr. and F.~Wilczek,
  ``On geometric entropy,''
Phys.\ Lett.\ B {\bf 333}, 55 (1994).
[hep-th/9401072].
}

\lref\BombelliRW{
  L.~Bombelli, R.~K.~Koul, J.~Lee and R.~D.~Sorkin,
  ``A Quantum Source of Entropy for Black Holes,''
Phys.\ Rev.\ D {\bf 34}, 373 (1986)..
}

\lref\PolchinskiTA{
  J.~Polchinski,
  ``String theory and black hole complementarity,''
In *Los Angeles 1995, Future perspectives in string theory* 417-426.
[hep-th/9507094].
}

\lref\MaldacenaDS{
  J.~M.~Maldacena and L.~Susskind,
  ``D-branes and fat black holes,''
Nucl.\ Phys.\ B {\bf 475}, 679 (1996).
[hep-th/9604042].
}

\lref\CzechBH{
  B.~Czech, J.~L.~Karczmarek, F.~Nogueira and M.~Van Raamsdonk,
Class.\ Quant.\ Grav.\  {\bf 29}, 155009 (2012).
[arXiv:1204.1330 [hep-th]].
}
\lref\HubenyWA{
  V.~E.~Hubeny and M.~Rangamani,
JHEP {\bf 1206}, 114 (2012).
[arXiv:1204.1698 [hep-th]].
}

\lref\BoussoSJ{
  R.~Bousso, S.~Leichenauer and V.~Rosenhaus,
Phys.\ Rev.\ D {\bf 86}, 046009 (2012).
[arXiv:1203.6619 [hep-th]].
}

\lref\NozakiWIA{
  M.~Nozaki, T.~Numasawa and T.~Takayanagi,
  ``Holographic Local Quenches and Entanglement Density,''
[arXiv:1302.5703 [hep-th]].
}

\lref\FaulknerYIA{
  T.~Faulkner,
[arXiv:1303.7221 [hep-th]].
}
\lref\HartmanMIA{
  T.~Hartman,
[arXiv:1303.6955 [hep-th]].
}

\lref\GibbonsUE{
  G.~W.~Gibbons, S.~W.~Hawking and ,
Phys.\ Rev.\ D {\bf 15}, 2752 (1977)..
}

\lref\DegerNM{
  S.~Deger, A.~Kaya, E.~Sezgin and P.~Sundell,
Nucl.\ Phys.\ B {\bf 536}, 110 (1998).
[hep-th/9804166].
}

\lref\MaldacenaBW{
  J.~M.~Maldacena and A.~Strominger,
JHEP {\bf 9812}, 005 (1998).
[hep-th/9804085].
}

\lref\WaldNT{
  R.~M.~Wald,
Phys.\ Rev.\ D {\bf 48}, 3427 (1993).
[gr-qc/9307038].
}

\lref\VilenkinZS{
  A.~Vilenkin,
Phys.\ Rev.\ D {\bf 23}, 852 (1981).
}

\lref\BoisseauBP{
  B.~Boisseau, C.~Charmousis and B.~Linet,
Phys.\ Rev.\ D {\bf 55}, 616 (1997).
[gr-qc/9607029].
}

\lref\IyerYS{
  V.~Iyer and R.~M.~Wald,
Phys.\ Rev.\ D {\bf 50}, 846 (1994).
[gr-qc/9403028].
}
\lref\IyerKG{
  V.~Iyer and R.~M.~Wald,
Phys.\ Rev.\ D {\bf 52}, 4430 (1995).
[gr-qc/9503052].
}

\lref\Igor{I. Klebanov, private communication. }

\lref\GubserBC{
  S.~S.~Gubser, I.~R.~Klebanov and A.~M.~Polyakov,
Phys.\ Lett.\ B {\bf 428}, 105 (1998).
[hep-th/9802109].
}

\lref\WittenQJ{
  E.~Witten,
Adv.\ Theor.\ Math.\ Phys.\  {\bf 2}, 253 (1998).
[hep-th/9802150].
}

\lref\FursaevIH{
  D.~V.~Fursaev,
JHEP {\bf 0609}, 018 (2006).
[hep-th/0606184].
}

\lref\NelsonNA{
  W.~Nelson,
Phys.\ Rev.\ D {\bf 50}, 7400 (1994).
[hep-th/9406011].
}

\lref\FaulknerPrivate{ T. Faulkner, private communication.}

\lref\FursaevEF{
  D.~V.~Fursaev and S.~N.~Solodukhin,
Phys.\ Rev.\ D {\bf 52}, 2133 (1995).
[hep-th/9501127].
}
\lref\SolodukhinGN{
  S.~N.~Solodukhin,
Living Rev.\ Rel.\  {\bf 14}, 8 (2011).
[arXiv:1104.3712 [hep-th]].
}

\lref\HungXB{
  L.~-Y.~Hung, R.~C.~Myers and M.~Smolkin,
JHEP {\bf 1104}, 025 (2011).
[arXiv:1101.5813 [hep-th]].
}
\lref\CasiniKV{
  H.~Casini, M.~Huerta and R.~C.~Myers,
JHEP {\bf 1105}, 036 (2011).
[arXiv:1102.0440 [hep-th]].
}
\lref\HungNU{
  L.~-Y.~Hung, R.~C.~Myers, M.~Smolkin and A.~Yale,
JHEP {\bf 1112}, 047 (2011).
[arXiv:1110.1084 [hep-th]].
}
\lref\SusskindSM{
  L.~Susskind and J.~Uglum,
Phys.\ Rev.\ D {\bf 50}, 2700 (1994).
[hep-th/9401070].
}
\lref\BanadosQP{
  M.~Banados, C.~Teitelboim and J.~Zanelli,
Phys.\ Rev.\ Lett.\  {\bf 72}, 957 (1994).
[gr-qc/9309026].
}
\lref\FursaevEA{
  D.~V.~Fursaev and S.~N.~Solodukhin,
Phys.\ Lett.\ B {\bf 365}, 51 (1996).
[hep-th/9412020].
}
\lref\CallanPY{
  C.~G.~Callan, Jr. and F.~Wilczek,
Phys.\ Lett.\ B {\bf 333}, 55 (1994).
[hep-th/9401072].
}
\lref\GubserPF{
  S.~S.~Gubser and A.~Nellore,
JHEP {\bf 0904}, 008 (2009).
[arXiv:0810.4554 [hep-th]].
}

\lref\BekensteinUR{
  J.~D.~Bekenstein,
Phys.\ Rev.\ D {\bf 7}, 2333 (1973)..
}
\lref\BardeenGS{
  J.~M.~Bardeen, B.~Carter and S.~W.~Hawking,
Commun.\ Math.\ Phys.\  {\bf 31}, 161 (1973)..
}
\lref\HawkingSW{
  S.~W.~Hawking,
Commun.\ Math.\ Phys.\  {\bf 43}, 199 (1975), [Erratum-ibid.\  {\bf 46}, 206 (1976)]..
}
\lref\FursaevSG{
  D.~V.~Fursaev,
Phys.\ Rev.\ D {\bf 77}, 124002 (2008).
[arXiv:0711.1221 [hep-th]].
}
\lref\FursaevMP{
  D.~V.~Fursaev,
JHEP {\bf 1205}, 080 (2012).
[arXiv:1201.1702 [hep-th]].
}



\Title{
\vbox{\baselineskip12pt
}}
{\vbox{\centerline{ Generalized  gravitational  entropy }\vskip .5cm
 \centerline{    }
}}

\bigskip
\centerline{ Aitor Lewkowycz$^1$ and Juan Maldacena$^2$ }
\bigskip

\centerline{ \it $^1$ Department of Physics, Princeton University,  Princeton, NJ, USA}
\bigskip
\centerline{ \it$^2$  School of Natural Sciences, Institute for
Advanced Study,Princeton, NJ, USA }

\vskip .3in \noindent

We consider classical Euclidean gravity solutions with a boundary.
The boundary contains a non-contractible circle. These solutions can be
interpreted as computing the trace of a density matrix in the full
quantum gravity theory, in the classical approximation.
When the circle is contractible in the bulk,
we argue that the entropy of this density matrix is given by the area
of a minimal surface. This is a generalization of the usual black hole entropy
formula to euclidean solutions without a Killing vector.

A particular example of this set up appears in the computation of the entanglement
entropy of a subregion of a field theory with a gravity dual. In this context,
the minimal area prescription was proposed by Ryu and Takayanagi. Our arguments
explain their conjecture.


 \Date{ }




\newsec{ Introduction }

Originally the concept of entropy arose from equilibrium thermodynamics.
However, we know think of entropy as a measure of information.
In particular, we can assign an entropy to a general density matrix via
\eqn\entrof{
S = - Tr[ \rho \log \rho ]
}

By thinking about the thermodynamics of black holes the area formula
for gravitational entropy was discovered \refs{\BekensteinUR,\BardeenGS,\HawkingSW}.
Gibbons and Hawking introduced a thermodynamic interpretation of
euclidean gravity solutions with a $U(1)$ isometry \GibbonsUE .
The idea is that one considers Euclidean solutions with prescribed
boundary conditions. The boundary conditions, as well as the solutions,
are invariant under a $U(1)$ symmetry\foot{Here we assume that there is a single $U(1)$ symmetry, otherwise
we need to add  the corresponding chemical potentials, etc.}.
These solutions can be viewed as
describing the computation of the partition function of a quantum theory in
the classical approximation. In other words, one thinks of the Euclidean
gravitational action as $\log Z(\beta) = - S_{E,grav}  $.
Then the entropy, obtained as $ S = - ( \beta \partial_\beta -1) \log Z $, is
equal to the area of the codimension two surface which is a fixed point for
the $U(1)$ symmetry in the bulk. Classically,
the boundary can be chosen to be any surface where we put boundary
conditions. It can also be an asymptotic boundary such as the $AdS$ boundary.

\ifig\densitymatrices{ (a) A  euclidean solution with a $U(1)$ symmetry is interpreted as computing the equilibrium
thermodynamic partition function of the gravity theory. (b) We consider a euclidean solution with a circle but
without a $U(1)$ symmetry. This is interpreted as computing $Tr[ \rho ]$ for an un-normalized density matrix in
the gravity theory. This is the density matrix produced by euclidean evolution. } {\epsfxsize3.5in\epsfbox{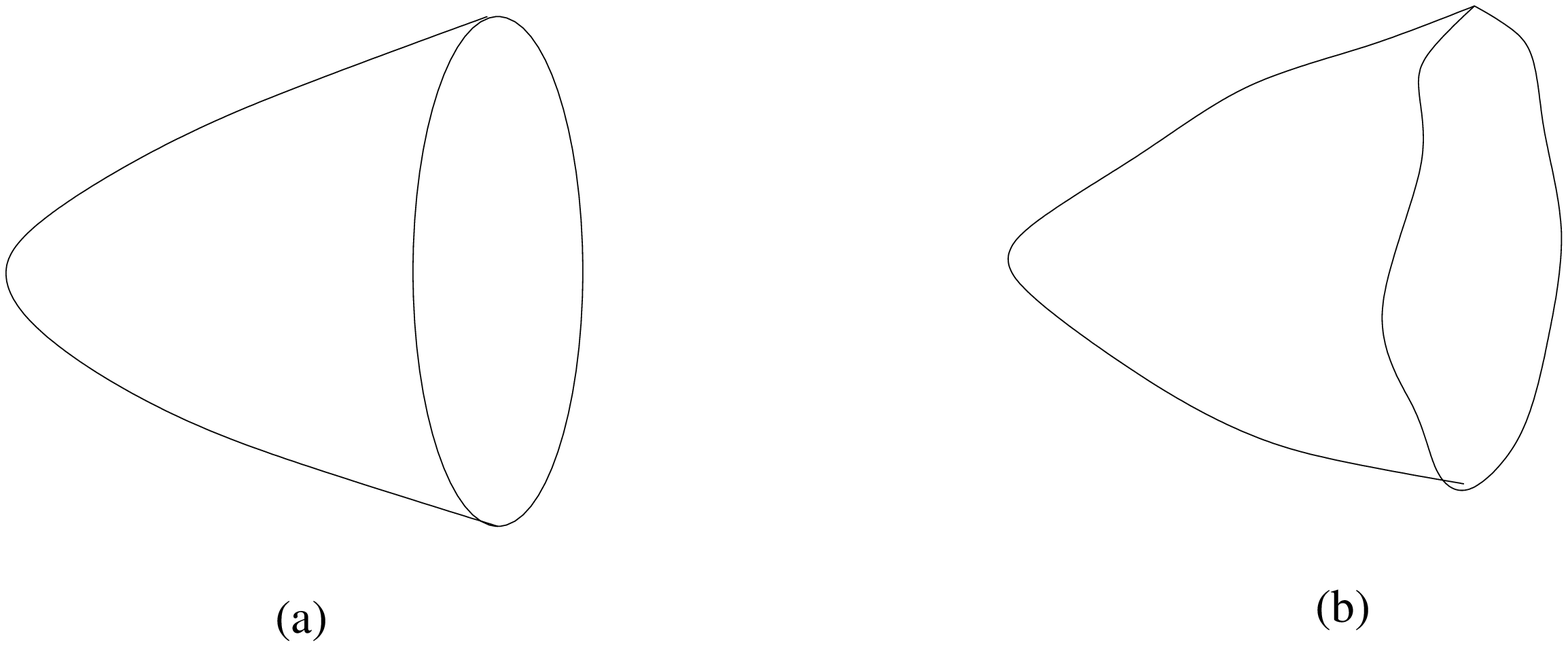}}

Interestingly, one can extend the notion of gravitational entropy to situations without a $U(1)$ symmetry as follows.

Let us first consider a general quantum system. Its Euclidean evolution generates
an un-normalized
 density matrix
\eqn\evolu{
 \rho  =   P e^{ - \int_{\tau_0}^{\tau_f} d\tau H(\tau ) }
 }
 where we considered a general time dependent Euclidean Hamiltonian.
We can compute the entropy of this density matrix by the ``replica trick''.
Namely, first notice that $Tr[\rho] $ can be computed  by considering euclidean evolution on a circle,
identifying $\tau_f =  \tau_0 + 2 \pi$ \foot{Throughout this paper we set the coordinate length of the initial circle
to $2\pi$. Of course, its physical length depends on the metric. }.  Similarly, we can compute $Tr[ \rho^n] $ by considering
time evolution over a circle of $n$ times the length of the original one, where the couplings in
the theory are strictly periodic under shifts of the original circle, $H(\tau + 2\pi) = H(\tau)$.

We then can compute the entropy as
\eqn\entrog{ \eqalign{
 S = &  - \left.   n \partial_n \left[ \log Z(n) - n \log Z(1) \right]\right|_{n=1}   = - Tr[ \hat \rho \log \hat \rho ]  ~,
 \cr
 & Z(n) \equiv Tr[ \rho^n] ~,~~~~~~~~~~~~ \hat \rho \equiv { \rho \over Tr[ \rho ] }
 }}
 where now $\hat \rho$ is a properly normalized density matrix. This involves computing $Z(n)$ and then performing
 an analytic continuation in $n$.

\ifig\Replica{ Computing the entropy using the replica trick. (a) Euclidean solution for $n=1$.
(b) Solution for $n=4$. At the boundary we go around the original circle $n$ times before making the identification.
We then find a smooth gravity solution with these boundary conditions. The curves in the right hand side are schematically
 giving the boundary conditions at infinity. We see that in (b), we simply repeat $n$ times the boundary conditions we had
 in (a). } {\epsfxsize3.5in\epsfbox{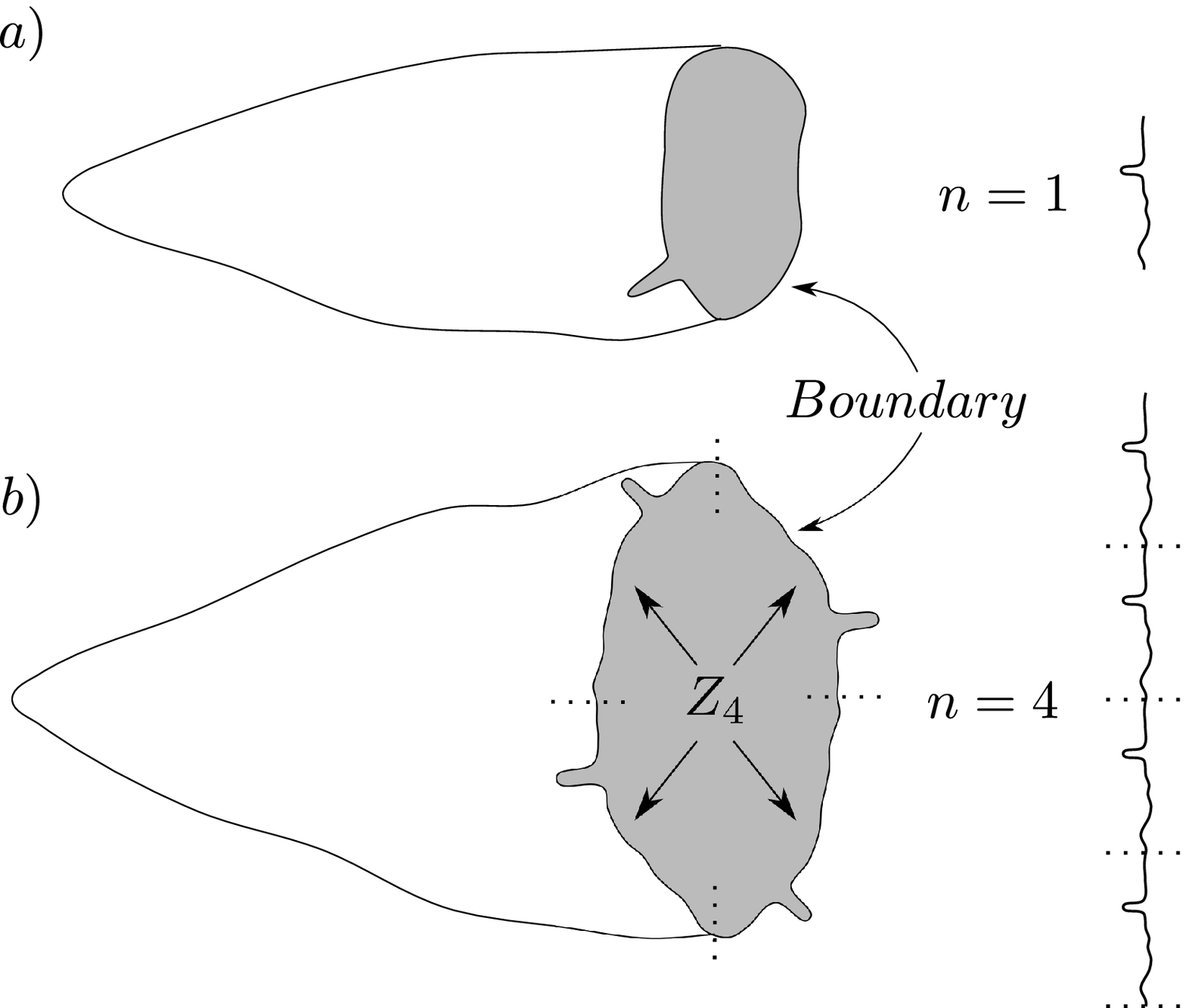}}

 Going back to the gravitational context, we can consider metrics which end on a boundary.
 We assume that the boundary has a direction with the topology of the circle. The boundary data can depend
 on the position along this circle but it respects the periodicity of the circle. We define
  the coordinate $\tau \sim \tau + 2 \pi$ on the circle.
   We can then consider a spacetime
 in the interior which is smooth. Its Euclidean action is defined to be $\log Z(1)$. See \densitymatrices (b).
  We can also consider other spacetimes where we take the same boundary data but consider
 a new circle with period $\tau \sim \tau + 2 \pi n $. Their Euclidean action is defined to be $\log Z(n)$.
 These computations can be viewed as computing $Tr[\rho^n]$ for the density matrix   produced by the
 Euclidean evolution.  See \Replica .
 If we are sufficiently diligent, we can find these actions, analytically continue in $n$ the
 corresponding answers and compute $S$ as in \entrog . This has been explicitly done in   \refs{\FaulknerYIA,\HartmanMIA} for
 some examples in three dimensional gravity.

Note that we are implicitly assuming that gravity is holographic. We are imagining that setting boundary
conditions on some boundary defines the theory and that the interior geometry is an  approximation
to the full computation. We do not know how (and whether) holography works for general boundaries. Here we only
need it to be approximately valid so that this classical computation has the interpretation of computing an approximate density
matrix in some approximate theory.
In cases where the boundary is a true asymptotic boundary (such as a locally asymptotically $AdS$ boundary)
the situation is well understood.
 This corresponds to computing the entropy of a perfectly well defined density matrix in
the dual field theory .

 Interestingly, there is a simple conjecture for the final answer.
 The entropy  is also given by the area of a special codimension two surface in the bulk of the
 original ($n=1$) solution.  At this surface
 the circle shrinks smoothly to zero size. The surface obeys a minimal area condition.
 \eqn\RTconj{
  S \equiv   - \left. n \partial_n \left[ \log Z(n) - n \log Z(1) \right]\right|_{n=1} =  { A_{\rm minimal}  \over 4 G_N}       ~~~~~~~~~~~~~~~
  }
  From now on, $\log Z(n)$ denotes the classical gravity action $\log Z(n) = -S_{\rm Grav}$ of the $n^{th}$ solution.
  This formula was first conjectured by Ryu and Takayanagi in the context of the computation of entanglement entropy
 of conformal field theories with gravity duals \RyuBV (see \NishiokaUN\ for a review\foot{
 See  \refs{\FursaevSG,\FursaevIX} for related work. }).
  Proving their formula amounts to proving the above conjecture,
 as we explain below.
 Notice that \RTconj\ can be viewed as
  a statement about classical general relativity. It is a relation between the actions for
  classical solutions  that are produced by the replica trick and the area of the minimal area
 solution with $n=1$. Of course, for solutions with a $U(1)$ symmetry, \RTconj\ reduces to the standard Gibbons-Hawking
 computation. In that case,  the $U(1)$ symmetry also ensures that the horizon is a minimal surface, with zero extrinsic
 curvature.

In this paper we will give an argument for \RTconj\  based on reasonable assumptions
regarding the analytic continuation of the solutions away from integer values of  $n$.

We will also explain why proving \RTconj\ is equivalent to proving the Ryu Takayanagi conjecture.
The Ryu-Takayanagi conjecture for the case of asymptotically $AdS_3$ pure gravity was proven in
\refs{\FaulknerYIA,\HartmanMIA}. Previous arguments   include \FursaevIH , whose
assumptions were criticized in   \HeadrickZT .\foot{
 Fursaev  \FursaevIH\ took the solution for $n=1$ and set   $\tau \sim \tau + 2 \pi n$ everywhere in the bulk.
This introduces a conical singularity in the bulk.
 As noted  by Headrick  \HeadrickZT , for integer $n$,
  one should instead consider solutions which are non-singular in the bulk.
  }

This paper is organized as follows. In section 2 we perform some explicit computations in a simple example.
In section 3 we review the derivation of the entropy formula for the case with a $U(1)$ symmetry.
In section 4 we present the arguments for the main formula \RTconj . There we explain how the solution looks for
$n$ close to one.  We also derive the minimal area condition for the surface.
In section 5 we discuss the connection to entanglement entropy in field theories with gravity dual.
In section 6 we present the conclusions. In the appendices we present some further explicit examples and more details
on the computations.

\newsec{ A simple example without a $U(1)$ symmetry }

Since our discussion has been a bit abstract, let us discuss a very simple
concrete example. This example will also motivate some assumptions that we will make
later.

Let us start with the BTZ geometry
\eqn\btz{
ds^2 =  \left[  { d r^2 \over (1 + r^2 ) } + r^2 d\tau^2 + (1 + r^2 ) dx^2  \right]
}
This metric has a $U(1)$ isometry along the circle labeled by $\tau$,  $\tau \sim \tau + 2 \pi $.
All functions will be invariant under translations in $x$. This direction will not play any
role in this discussion and we take it to be compact of size $L_x$.
 Computing the entropy for this solution gives the standard area formula, $S_0$, for this
solution.

We now add a complex, minimally coupled,  massless scalar field $\phi$.
We  set boundary conditions  that
are not $U(1)$ invariant
\eqn\bcsp{
\phi = \eta  e^{ i \tau } ~,~~~~~~~~~~{\rm at } ~~~~ r =\infty
}

We now compute the gravitational action to second order in $\eta $ for the family of solutions
described above. The metric is changed at order $\eta^2$, but since the original background obeys Einstein's equations,
there is no contribution from the gravitational term to order $\eta^2$.   So, to this order, the whole contribution comes
from the scalar field term in the action.

Namely, for the $n^{th}$ case, we need to consider a spacetime with the same boundary conditions as in
\bcsp\ but where $\tau \sim \tau + 2 \pi n $. This implies that the spacetime in the interior is
\eqn\intb{
ds^2 =  \left[  { d r^2 \over (n^{-2} + r^2 ) } + r^2 d\tau^2 + (n^{-2} + r^2 ) dx^2  \right]
}
And we need to consider a scalar field in this spacetime.
We can write the wave equation.
 The solution of the wave equation that is regular at the origin and obeys  \bcsp\ at infinity  is
\eqn\solwv{
\phi = \eta e^{ i \tau } f_n( r ) ~,~~~~~~ ~~~~~~~~f_n(r) = (n r)^n { \Gamma( { n \over 2 } +1 )^2 \over      \Gamma(n+1)} \, \,
 _2 F_1\left(\frac{n}{2} ,\frac{n}{2}+1  ;n+1;-( n r)^2\right)
}
Note that $f_n \to 1 $ as $r \to \infty $.

We now evaluate the gravitational action for every $n$. We evaluate it to second order in $\eta$, so we consider
 the quadratic action for the field $\phi$.
Using standard formulas we can write
\eqn\sacc{\eqalign{
\left. \log Z(n)\right|_{\eta^2} & =-  \int_{AdS_3}  | \nabla \phi |^2  = - (2 \pi n ) L_x  \left[  r^3 \phi^{*} \partial_r \phi \right]_{r = \infty}  =
\cr
 & = ( 2 \pi L_x) \left[  1 - n \log n + n \psi(n/2)  + ( {\rm linear~in~} n ) \right]
 }}
where $L_x$ is the length of the $x$ direction and $\psi$ is the Euler $\psi$ function.
 The terms linear in $n$ include divergent terms that should be subtracted.
However, they do not contribute to the entropy \entrog .

We  analytically continue in $n$ and compute the entropy via \entrog\  to find
\eqn\finc{
S = S_0 + \eta^2  \pi L_x  ( 4 - { \pi^2 \over 2 } )
}

We can now compare this with the answer we expect from the area formula.  This non-zero configuration for
the scalar field changes the geometry to second order in $\eta$. Thus it produces a second order change in the
area of the horizon.  This change can be computed from
Einstein's equation. We obtain the same answer \finc . This is done in detail  in appendix A, where we also consider
a scalar field with an arbitrary mass.

So, we have explicitly checked the conjecture for this special case. Now, let us make some remarks.

We considered a complex scalar field, but the computation can be done also for a real scalar field with
boundary conditions $\phi = \eta \cos \tau $ at infinity. The result is essentially the same. See appendix A.

Notice that the solution for the $n^{th}$ case has a $Z_n$ symmetry. This is a replica symmetry of
the boundary conditions which extends to the bulk solution. So, in this case we are not breaking
the replica symmetry. Notice that $r=0$ is a fixed point of the action of the $Z_n$ replica symmetry for
all $n > 1$. In this case, the metric has a $U(1)$ symmetry. However, the full scalar field
configuration is only symmetric
under the $Z_n$\foot{ In this case there is a $U(1)$ symmetry which is shift in $\tau$ combined with a phase rotation
of the complex field. But for a similar computation with a real scalar field we only have the  $Z_n$ symmetry.}.

Here we have computed $\log Z(n)$ and then analytically continued the answer.
 The geometry \intb\ is well defined also for non-integer $n$ and
we can trivially continue it to non-integer values of $n$, and it remains smooth.
We could ask whether we can also
 analytically continue the whole field configuration to  non integer values of $n$.
Notice that as we vary $n$, the $\tau$ dependence at the boundary is kept fixed. Thus, even for non-integer $n$,
we will keep the same boundary condition. This boundary condition is not compatible with a non-integer period
for $\tau$. We will ignore this. In other words, we will integrate $\tau$ between $[0,2\pi]$ and multiply the result by $n$.
However, as we go to $r=0$, we find that the scalar field behaves as $\phi \sim r^n e^{ i \tau } $,
which leads to a singularity for the scalar field at $r=0$. The scalar field, or its stress tensor do not
diverge if $n>1$. In other words, this appears to be a relatively harmless integrable singularity.
 This singularity seems physically questionable.  But we are not trying to give a physical interpretation to the
 solution with non-integer $n$. We are only trying to define it mathematically, as an intermediate step in
 computing the replica trick answer. One could worry that if we allow singularities, then the solution will not be
 uniquely defined. However, we are allowing a very specific behavior which determines a unique solution for given
 boundary conditions. More explicitly, note that when we solve the wave equation near $r\sim 0$, we get two
 solutions $ r^n e^{ i \tau} $ and $r^{-n} e^{ i \tau }$. We set to zero the coefficient of the second solution at
 the origin. This  prescription uniquely selects a solution, both for integer and non-integer $n$.

This gravity  theory in $AdS_3$ with a massless scalar field can arise  from a
Kaluza Klein reduction of a higher dimensional theory. For example, it can come from a ten dimensional
solution of the form
$AdS_3 \times S^3 \times T^4$. Then the massless field can be an off-diagonal component of the
metric on of the four torus \refs{\MaldacenaBW,\DegerNM}.
 More explicitly,  we can deform the metric of the four torus as
 \eqn\metrcto{
 ds^2_{  T^4 } = e^{2 \phi_1} dy_1^2+e^{2 \phi_2} dy_2^2+e^{-2 \phi_1} dy_3^2+e^{-2\phi_2} dy_4^2
 }
 where $\phi = \phi_1 + i \phi_2$.
  We   see that
 the singularity of the field $\phi$ at the origin translates into a singularity for some of the Riemann tensor components.
 Let us consider $n = 1 + \epsilon$. Then since  $\phi \propto r^{ 1 + \epsilon } e^{ i \tau} $ this leads to
 a singularity in some of the Riemann tensor components  $R_{ \alpha  i \beta i} \sim {\epsilon \over r}$  (no sum over $i$), where $i$ are the directions on the four torus and $\alpha $ denotes the directions along the two transverse components (labeled also
 by $r$ and $\tau$)
 Despite these singularities the action is finite, as we saw when we computed it explicitly.

An alternative way to view the solution labeled by $n$ is the following.
 We consider the $\tau$ circle to have period $2\pi$  but introduce a
conical singularity at the origin with opening angle $2\pi/n$. This is not the same as the gravity solution with
$n=1$ since the field configuration has to adjust to the presence of the conical singularity.
Then, when we evaluate the gravitational
action, we integrate  $\tau$ over $[0,2\pi]$  but multiply the resulting answer by a factor of $n$.
 This factor of $n$ arises because the real period of $\tau$ is $2\pi n$ instead of $2\pi$. It is important
 that we evaluate the gravitational action
{\it without introducing any contributions from the tip of the conical singularity}, since the full space
(with the right period for $\tau$) is non-singular  \foot{This looks superficially
similar to what was discussed in  \FursaevIH, but it is different in detail.  } \foot{
 This point of view was also  suggested to us by T. Faulkner.} . This picture makes sense both
 for integer or non-integer $n$.

 \newsec{ Computation of gravitational entropy when there is a $U(1)$ symmetry }

In this section we will describe the computation of the entropy using Euclidean methods in a way
that it emphasizes the fact that the contribution comes form the horizon. This has been discussed by various
authors in a similar form
\refs{\BardeenGS,\CarlipSA,\BanadosQP,\SusskindSM,\FursaevEF,\NelsonNA}.
 Here we say it two  ways  that we particularly liked.

\subsec{ Entropy from rounded off cones }

 \ifig\fourcones{ A particular combination of geometries that is useful for computing the entropy.
 The first geometry is the correct  solution with period $2\pi n$.
 The last geometry contains a conical singularity. It is the solution with $n=1$ but with the circle identified after
 $\tau \to \tau + 2 \pi n$. For $n=1$ the deficit angle of the cone is very small and it has been greatly exaggerated here
  for artistic reasons.
  The two middle ones are identical and correspond to a regularized version of the last solution.
 They only differ for $r< a$, where $a$ is small regulator. This is not a solution, it is an off-shell configuration. All of the configurations obey the same boundary conditions at infinity. } {\epsfxsize3.5in\epsfbox{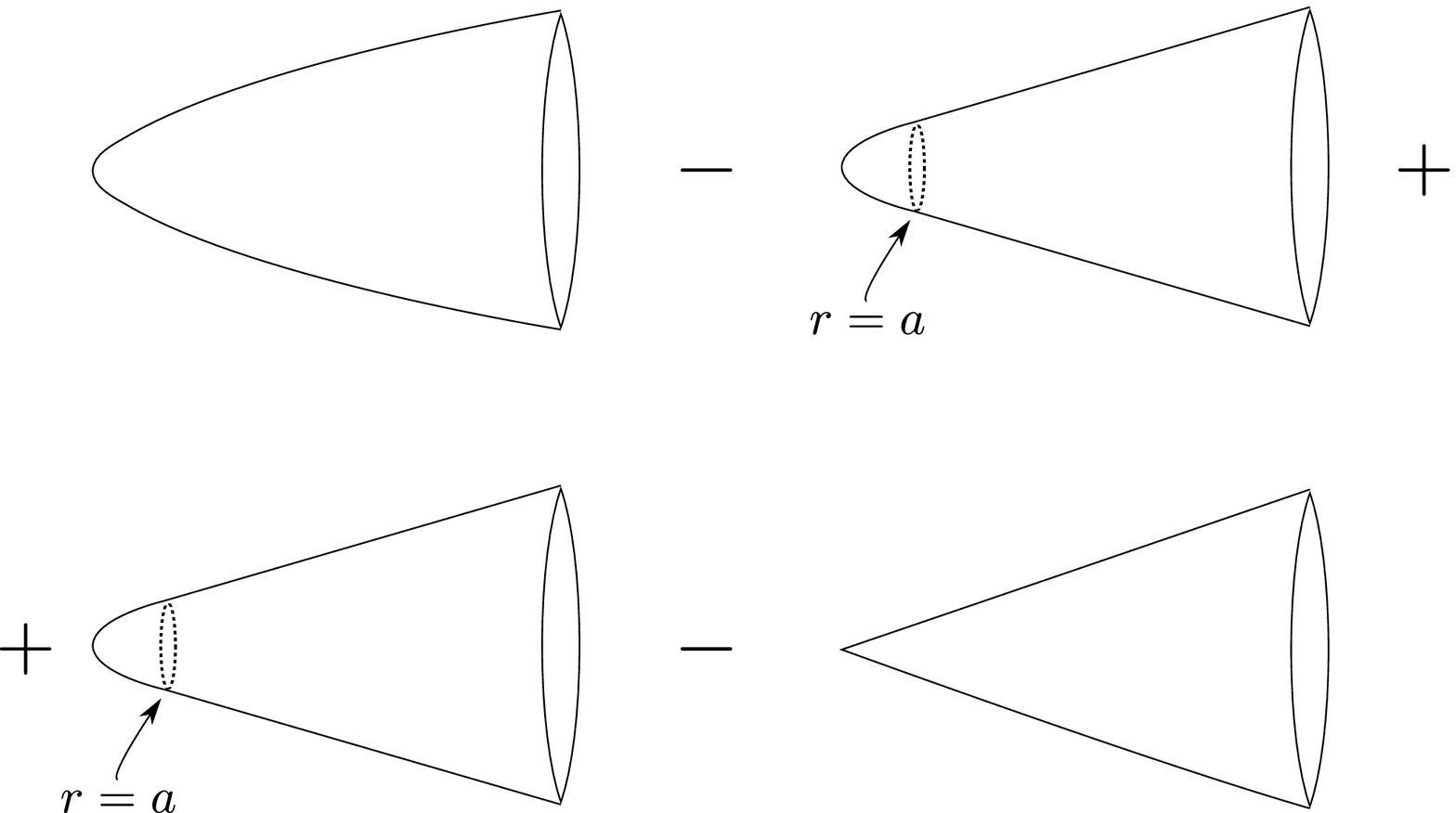}}

Setting the period of the circle to be $ \tau \sim \tau + 2 \pi n $, then we find that the formula for the entropy
can be written as
\eqn\etnr{
  S =  - n  \partial_n \left[ \log Z(n) - n \log Z(1)\right]_{n=1}
  }
  Let us consider this expression for $n$ close to one.
  We interpret the first term in the square brackets   as the correct, smooth solution when $n$ is not one.
  We interpret the second term as the solution for $n=1$ but with a $\tau$ which has period $\tau \sim \tau + 2 \pi n$.
  This solution has a conical singularity at the origin.
   However, we {\it do not } include any contribution from the conical singularity.
  We simply integrate the gravitational action density away from the tip.

  We now evaluate the difference in the square brackets in \etnr\
  by adding and subtracting a smooth geometry which is the same
  as that of the cone far  away from the origin, but it is a regularized cone near the origin, see \fourcones .
  This smooth geometry is not a solution of the equations of motion of the theory, it is an off-shell configuration.
   We are simply introducing it to help us perform the computation. It is possible  to choose this off-shell configuration
   in such a way that the metric differs only by an amount of order $n-1$ from the true solution.

  Thus we get
 \eqn\etnr{
  S =  -   n \partial_n \left[ ( \log Z(n)  - \log Z^{\rm off}(n) ) + ( \log Z^{\rm off}(n) - n \log Z(1) ) \right]_{n=1}
  }
  Each of the terms in the brackets is the action for one of the configurations in \fourcones .
  Since the off shell configuration that corresponds to a regularized cone differs by a first order term in $n-1$ from
  a solution of the equations, we see that we can interpret the first parenthesis as the result of doing a first order
  variation away from a solution (the solution with period $n$). This first order variation vanishes due to the equations
  of motion for the solution with period $n$. Notice that both metrics obey the same boundary conditions at the
  boundary, so that there are no boundary terms\foot{The absence of boundary terms is clearest if we write the action
  in a non-manifestly covariant form using only first derivatives of the metric.
   Then the fact that the two configurations obey
  the same boundary conditions for the metric implies that there are no boundary terms.}.

  So all that remains is the second parenthesis. The second parenthesis contains the difference between a smooth cone
  and a regularized cone. This receives a contribution only from the region near the tip of the cone.
  This contribution
  is extensive in the area of the horizon, namely the area of the surface transverse to the tip of the cone.
   The region near the rounded  tip of the cone
   contains an integral of $\int d^2 x \sqrt{g} R$ along the cone directions which gives
  \eqn\inttwod{
  \int_{\rm Reg~Cone}  d^2x \sqrt{g} R \sim 4 \pi ( 1 -n )
  }
  Thus, the final answer has the form
  \eqn\finans{
  S = { 1 \over 16 \pi G_N } ({\rm Area} ) \left( - n \partial_n  \int_{\rm Reg~Cone}  d^2x \sqrt{g} R \right)   = { { \rm Area } \over 4 G_N }
  }

 One can consider a metric that explicitly regularizes the cone, such as \FursaevEF
 \eqn\metre{
  ds^2  =  d r^2 g^2(r) + r^2 d\tau^2
}
where $g = n + o(r^2)$ at $r\sim 0$ and $g =1$ for $r > a$, where $a$ is a small distance which sets the size of the
regularization. Inserting this metric into the gravitational action we get  \inttwod . One can choose a completely
explicit function such as $ g = 1 + (n-1) e^{ - r^2/a^2 } $, for example. In this case we can see explicitly that
the metric perturbation   is of order $(n-1)$.

\subsec{ Entropy from apparent  conical singularities}

Another way to think about this problem is as follows.
First we note that, since the solutions are invariant under time translation, the evaluation of
$\log Z(n)$ is the same as
\eqn\evallg{
\log Z(n) =   n  [ \log Z(n)]_{2\pi}
}
where $ [ \log Z(n)]_{2\pi} $ is the gravitational action density for the solution labeled by $n$ but
integrated over $\tau$ from $[0,2\pi]$ (instead of $[0,2 \pi n]$).
 We can now write the entropy as
\eqn\entrp{
S =-   n^2  \partial_n  [ \log Z(n ) ]_{2\pi}
}
Note that the solution labeled by $n$ is a smooth geometry if the $\tau $ circle has period
$ 2 \pi n $. On the other hand, imagine we wanted to view it as a configuration where
the $\tau $ period continues to be $2\pi$. In that case, it is a geometry with a conical singularity
whose opening angle is $2\pi/n$.  Thus we can view
\eqn\func{
[ \log Z(n ) ]_{2\pi}
}
as the gravitational action of a configuration with $\tau = \tau + 2 \pi$ but with a conical
singularity with opening angle $2\pi/n$, {\it without including any curvature contribution from the
conical singularity}.
Then we see that the expression of the entropy \entrp\ involves taking a derivative with respect to
$n$. When we change $n$ we are changing the opening angle of the singularity. In addition,  we are
changing the metric and other fields everywhere since they have to adjust to this new strength of the
singularity. However, since the original solution (the solution with $n=1$) is a solution of the equations,
we would naively expect that a first order variation of the metric and other fields should vanish due to their
 equations of
motion. This naive expectation is essentially right, except for the fact that we are changing the boundary
conditions at the origin, since the strength of the conical singularity is being changed.
Thus, the only change in the action comes from a boundary term.
In other words, when we change $n$ the action
changes as
\eqn\changec{\eqalign{
  -\partial_n  [\log Z(n)]_{2 \pi }  |_{n=1} = &   \int  E_g \partial_n g + E_\phi \partial_n \phi  +
  \cr &
  + {1 \over 8  G_N} \int_{r\sim 0} dy^{D-2} \sqrt{g}
   ( \nabla^\mu \partial_n g_{\mu r} - g^{\mu\nu} \nabla_r \partial_n g_{\mu \nu} )=
{A \over 4 G_N}
}}
where $E_g$ and $E_\phi$ are the equations of motion for the metric and other fields, which vanish. Here $y$ are
the coordinates along the $r=0$ surface.
 The boundary term vanishes at the large
$r$ boundary since we are choosing boundary conditions in such a way that the variation of the action
gives the equations of motion without extra boundary terms. On the other hand, at the horizon (at $r=0$),
we do get a contribution from the boundary term. This boundary term produces the area contribution. Note
that the $n$ derivatives of the metric are evaluated at the horizon. For example, in the
parametrization $ds^2 = n^2 dr^2 + r^2 d\tau^2$ near the origin, we get, as the only non-vanishing
component,  $\partial_n g_{rr}|_{n=1} = 2 $. With these expressions we can evaluate the parenthesis in \changec\
and obtain $2/r$.

This derivation easily generalizes to theories with higher derivative actions, giving
the Wald entropy \refs{\WaldNT,\IyerYS,\IyerKG}.

Note that in both cases we used explicitly the locality of the action along the $\tau$ direction. It would
be interesting to find the corresponding formula in weakly coupled string theory  exactly in $\alpha'$.

\newsec{ Argument for  the entropy formula $\RTconj $  }

\subsec{ Properties of the metric for $n$ integer }

For $n=1$,
the boundary contains  a  circle which we  label by the coordinate $\tau$. Recall that the boundary is the surface
where we are putting boundary conditions. This circle is non-contractible on the boundary, but it can be contractible in
the interior of the geometry. Here by boundary, we mean the boundary where we set boundary conditions for the
gravitational action. It need not be an asymptotic boundary.

The metric and all fields are periodic on this circle. Let us collectively denote
these fields as $\psi(\tau)$, with
\eqn\perfields{
\psi(\tau) \sim \psi(\tau + 2 \pi )
}
Of course, the fields depend on other coordinates, but
here we are highlighting their $\tau$ dependence.
We impose boundary conditions
\eqn\bcalf{
\psi(\tau)|_{\rm Boundary} = \hat \psi_B(\tau) ~,~~~~~~~~~~\hat\psi_B(\tau) = \hat \psi_B(\tau + 2 \pi )
}
where we specify the functions $\hat \psi_B(\tau)$, which are periodic.

The solution with $n> 1$, has exactly the same boundary conditions \bcalf , but
we require the periodicity $\tau = \tau + 2 \pi n$ on the $\tau $ circle.
This implies that the boundary conditions have a $Z_n$ symmetry.
We assume that the bulk solution continues to have this $Z_n$ symmetry.\foot{
In principle, the replica
$Z_n$ symmetry can be broken. Our discussion assumes that it is not broken.
The simplest gravity solutions can also develop other instabilities.
 For example, if one considers gravity in $AdS_{d+1}$ with
a boundary $H_{d-1} \times S^1$. If the radius of $S^1$ is  equal to the radius of $H_{d-1}$ then the full solution
is $AdS$, viewed as a black brane with a hyperbolic spatial section. If we make the $S^1$ $n$ times  larger, then
for large $n$, we approach an extremal black hole with an $AdS_2 \times H_{d-1}$ near horizon geometry. This can lead
to bad tachyons for $m^2 R_{AdS_{d+1}}^2 < - d/4$. Thus if the original $AdS_{d+1}$ has tachyons in the allowed
range  $ - d^2/4 \leq  m^2 R_{AdS}^2 <  -d/4 $, then we will have an instability.
This is similar to the discussion of \GubserPF , where  an extremal  Reissner-Nordstrom black brane
in $AdS_4$, with a near  horizon geometry  $AdS_2 \times R^2$ was considered. Good tachyons in $AdS_4$ can be bad tachyons in the $AdS_2$ region if $-{9 \over 4} <m^2 R_{AdS}^2<-{3 \over2}$. See
\BelinDVA\ for further discussion. Here we assume that we have no dangerous tachyons that can
lead to these instabilities.  Similar instabilities were   observed   computing the Renyi entropies of
circular regions in the three dimensional interacting $O(N)$ model \Igor .
 }.

Each of the solutions for $n>1$ has a special codimension two
surface which is left invariant by the action of $Z_n$. We will
focus on this surface. We can choose a coordinate $r$ which is a radial coordinate away from this surface and
an angle $\tau$. The true angle around the surface is really $\alpha = \tau/n $, we have chosen $\tau$ to have
the same period as the one we have at infinity ($\alpha \sim \alpha + 2 \pi $). The metric in the two directions transverse to
this surface has the form
\eqn\metric{
ds^2 = { n^2  dr^2   } + r^2 d\tau^2 + \cdots
}
where the factor of $n$ comes from demanding that there is no singularity at $r=0$.
 In addition,  all fields
are  required to have an $e^{ i k \tau }$ dependence, with integer $k$. This comes from the period of $\tau$  and
the $Z_n$ symmetry. Thus, a scalar field would behave as $r^n e^{ i \tau } \sim r^n e^{ i n \alpha } $ near the origin,
as results from demanding that it is non-singular.

As a side remark, notice that if the bulk space has no fixed points under the $Z_n$ action, then this means that
we can choose the coordinate $\tau$ in the interior so that this circle never shrinks.
 An example is a space with topology $R^{d-1} \times S^1$,
but with a metric that depends on the coordinate along the $S^1$.
In these cases the entropy is zero.
  The reason is very simple, the solution for the  $n^{th}$ replica is the same as the solution with $n=1$ but
with a longer circle so that $\log Z(n) = n \log Z(1)$. Here, of course, we used the locality of the classical action.

\subsec{ Metric for $n$ non-integer }

Here we make some assumptions on the form of the metric when $n$ is not an integer.
We will continue to impose exactly the same boundary condition \bcalf , which is periodic with
period $2 \pi $. This is {\it not} compatible with $\tau \to \tau + 2 \pi n$. Now, in the region
where the $\tau $ circle has positive size, we can ignore this problem and think of $\tau$ as being
non-compact. When we evaluate the action, we can integrate  the $\tau $ direction from 0 to $2\pi$ and
then multiply by $n$.

However, we expect that there is still a surface where the $\tau $ circle shrinks to zero. For the two
dimensions transverse to this surface  we
impose that the metric continues to behave as in \metric , even though $n$ is not an integer.
The rest of the fields, including other components of the metric, are chosen so that they are periodic in
$\tau \to \tau + 2 \pi $ as in \perfields.
 This implies that the field configuration is singular at $r=0$. However, we expect that this
singularity is as harmless as the one we had for the scalar field in section 2 .

This is seems a reasonable assumption. As evidence for its validity we can point to the
explicit example mentioned in section 2 .

An  equivalent way to specify the solutions is to compactify the $\tau$ circle to $\tau + 2 \pi $ in all
cases (all values of $n$) and  demand that there is a conical defect angle with opening angle $2\pi/n$ in the interior.
We do not introduce any contribution to the action from the tip of the cone.
In addition,  we   multiply the gravitational action by a factor of $n$.
 This is mathematically equivalent to
what was discussed above and the reader can choose the preferred interpretation.

Note that this is similar to introducing a cosmic string (or cosmic $D-3$ brane)
 with opening angle $2 \pi/n$ in the original
 solution, with the metric backreacting as necessary to account for its presence.

As $n\to 1$   the solution goes over to the solution with $n=1$.
Thus, this analytically continued solution is close  to the $n=1$ solution and we can expand it in
powers of $n-1$.

\subsec{ Derivation of the minimal surface condition}

We emphasized that for $n>1$ we have a special surface where the circle shrinks, and is fixed  under
 the $Z_n$ action. But for $n=1$ there is no obvious special surface,
 since there is no unique way to choose the coordinate $\tau$ in the
interior once it is not associated to a $U(1)$ isometry. So, when we expand the solution in $n-1$ we need to
select a surface. Motivated by the Ryu-Takayanagi conjecture we want to select a minimal area surface. In this subsection
we will explain the origin of this minimal area condition. The final conclusion is that the condition comes
from demanding that the solution obeys the Einstein equations to leading order in $n-1$.
This derivation is essentially the same as the derivation of the equations of motion for a cosmic string
(or $D-3$ brane) from the behavior of the metric near the conical singularity. This
problem was analyzed previously in  \refs{\UnruhHY,\BoisseauBP}.

{\it Two dimensional dilation gravity }

It is good to start with a simple situation first. For that purpose we will consider a two dimensional dilaton gravity
  where the action is
\eqn\twoddg{
-S_{\rm Grav} = { 1 \over 16 \pi } \int d^2x \sqrt{g}  e^{ - 2 \varphi} \left[   R +  4 ( \nabla \varphi)^2 + \cdots \right]
}
where the dots indicate other fields, or a potential for $\varphi$, etc. Notice that if we have a solution with a
horizon, then $e^{ - 2 \varphi} $ at the horizon
plays the role of the area in Planck units of the higher dimensional gravity solutions.
 In this case the codimension two surface is just a point.
The minimum  area condition is that $e^{ - 2 \varphi } $ is  a minimum (or really an extremum)
 at this point.
We will derive this condition from demanding that the configuration for  small $\epsilon \equiv n-1$ obeys the linearized
field equations near $r=0$.
In other words, expanding the fields around the $n=1$ solution, and {\it assuming the periodicity condition for the
fields}, \perfields ,  we will see that we can only obey the equations if $\partial_i \varphi =0$.

Let us say  that as $n \to 1 $ the special surface goes over to some point  of the $n=1$ manifold. Let us
pick this point to be the origin in some coordinate system  $x^1,x^2$. Then the metric of the $n=1$ solution around this
point is $ds^2 = dx_1^2 +dx_2^2 + o(x^2)$. The field $\varphi$ is regular at this point. Now, for $n-1 = \epsilon$ we
expect a metric of the form $ ds^2 = e^{ 2 \rho} ( dx_1^2 + dx_2^2 )$, with
$ e^{ 2 \rho } = r^{ 2 ( { 1 \over n } -1 ) } $, as $r \to 0$.
 Then to first order in $\epsilon $ we have
$\delta \rho \sim -  \epsilon  \log r $ to be the first order solution.
We consider the two following equations for two dimensional dilaton gravity
\eqn\dilgra{ \eqalign{
   0 = &  e^{ -2 \varphi } ( 4 \partial_z \varphi \partial_z \rho  +2 \partial_z^2  \varphi )  +   T_{zz}^{\rm matter}
   \cr
   0 = &  e^{- 2 \varphi } ( 4 \partial_{\bar z} \varphi  \partial_{\bar z} \rho+ 2\partial_{\bar z} ^2   \varphi )  + T_{\bar z \bar z}^{\rm matter}
 }}
 where $z = x^1 + i x^2$, $\bar z = x^1 - i x^2 $.
 Here $T^{\rm matter} $ denotes the stress tensor for the rest of the fields of the theory, coming from the dots in \twoddg . Expanding the first
 equation to first order we find
 \eqn\expfio{
 - 2 \partial_z \varphi(0)  { \epsilon \over z }  + 2 \partial_z^2  \delta \varphi     + \delta T_{zz}^{\rm matter } =0
 }
 and a similar equation by expanding the second. Here $\partial_z \varphi(0)$ is the derivative of the field for the
 $n=1$ solution at the origin. It is just a $z$ independent constant.
 Since the matter stress tensor is not expected to be  singular at order $1/r$,
 we find that
 \eqn\condc{
 \partial_z^2  \delta \varphi  \propto  { \epsilon \partial_z \varphi(0)  \over z } ~,~~~~~~~~~ \partial_{\bar z}^2 \delta \varphi \propto
 { \epsilon \partial_{\bar z } \varphi(0)  \over \bar z }
}
up to terms that are less singular as $r\to 0$.
Now we assume that the solution for $\delta \varphi$ has a fourier expansion with integer powers of $e^{ i \tau } $.
The first equation in \condc\ suggests that we try a solution proportional to $ \delta \varphi \propto z \log z$.
However, the periodicity condition under shifts of $\tau$ suggests that we should consider $\delta \varphi \propto
z \log( z \bar z )$. However, this is not a solution of the second equation. Thus this implies that the gradients of
the field should vanish at the origin.

More formally, we can argue as follows. The periodicity condition implies that
  if we take  the  $\tau$ derivative of any field  and integrate over $\tau$ between zero and $2\pi$, we
should get zero. This is true both for $\delta \varphi$ and its derivatives. In particular, note that the
following combination of derivatives gives
\eqn\comin{
 \partial_\tau \left[ ( r\partial_r - 1)  \partial_z \delta \varphi \right] \propto ( z \partial_z - \bar z \partial_{\bar z } ) (
 z \partial_z + \bar z \partial_{\bar z } -1) \partial_z \delta \varphi \propto \epsilon \partial_z \varphi(0)
 }
 where we used both equations in  \condc . Now the integral over $\tau$ of \comin\ should be zero, according to our assumption
 about the periodicity of $\delta \varphi$. This then implies that $\partial_z \varphi(0) =0 $.

In summary, in this case we found that the condition comes from the $zz$ and $\bar z \bar z$ components of the
 Einstein equations.
In higher dimensions we expect that this will come from Einstein's equations in the directions normal to the surface.

Note that if we changed the coefficient of the dilaton kinetic term in \twoddg\ from $4 (\nabla \varphi)^2 $ to
$ (4 + \sigma ) (\nabla \varphi)^2$, then we would be adding terms of the form $\sigma \partial_z \varphi
\partial_z \varphi$ to the equations in \dilgra . Expanding around the background solution such terms lead
to contributions that are subleading, in the expansion around the origin, compared to the terms already taken into
account in \condc . Thus, if we had a two dimensional action with a different coefficient for the dilaton kinetic
term, we would have reached the same conclusion\foot{This is to be expected since this coefficient can
be changed by a field redefinition of the metric.}.

{\it Einstein gravity in $D$ dimensions }

We now go back to the case of Einstein gravity.
In general, we can expand the metric of the $n=1$ solution  around the special surface as
\eqn\expme{\eqalign{
ds^2  = &  dr^2 + R^2 d\tau^2 + b_i d\tau dy^i +  g_{ij}  dy^i dy^j ~,~~~~~~~~~
\cr
g_{ij} = &
 h_{ij} +  r  \cos \tau K^1_{ij} + r \sin \tau K^2_{ij}  + o (r^2)
\cr
R = & r  + o(r^3)  ~,~~~~~~ b_i = o(r^2)
 }}
 where $r$ is coordinate normal to the surface and $y_i$ are coordinates along the surface. Here $K^\alpha_{ij}$ are
 the two extrinsic curvature tensors. $h_{ij}$ depends only on $y_i$ but not on $r$ or $\tau$.
 When we deform away from $n=1$ we assume that we cannot change the period of the cosines above.

When  $n=1+ \epsilon$   some of the metric components  generically  go like
$r^{ 1 + \epsilon}$. This  can give rise to terms in the equations of motion going like $1/r$. These terms can only
come from situations where we have two derivatives along the transverse directions (the $r$ and $\tau$ directions).
Such terms in the equations of motion are the same as the ones we would obtain by performing a dimensional reduction
from $D$ dimensions to the two transverse directions. This  brings us back to the previous case.
More explicitly, we write the full $D$ dimensional metric as
\eqn\tendm{
ds^2 =  
 e^{ 2 \rho } ( dx_1^2 + dx_2^2 )  + e^{ - {4 \varphi \over D-2}}
  \hat g_{ij} dy^i dy^j   + o (r^2) ~,~~~~~~~~~~~~\det( \hat g_{ij} ) =1
}
where $\hat g_{ij}$ is the transverse metric appearing in \expme\ but rescaled so that  its determinant is one.
The off diagonal terms in \expme\ do not contribute to terms of order $1/r$ in the equations of motion.
Both $\hat g_{ij}$ and $\varphi$ depend on all the coordinates, the $y_i$ and the $x_i$.
Here we have pulled out the overall volume factor of the transverse space and parametrized it by $\varphi$.
Dimensionally reducing to the first two dimensions gives us \twoddg, but with a different coefficient for the dilaton
kinetic term.
Thus, we obtain the same conditions that $\partial_{x^\alpha} \varphi =0$ for $\alpha =1,2$.
Now if we translate between $\varphi$ and the original metric \expme\ we find that
\eqn\mecg{
 - 4  \varphi  = \log( \det( h_{ij} ) )  +  x^1 K^1 + x^2 K^2 + o(r^2)  ~, ~~~~~~~~~~~~K^\alpha = h^{ij} K^\alpha_{ij}
 }
 where $K^\alpha$ are the traces of the extrinsic curvature tensors.
 We then see that the condition $\partial_\alpha \varphi =0$ implies that
 \eqn\msec{
 K^1 = K^2 =0
 }
 Namely, the traces of the extrinsic curvatures should vanish. There are two directions that are transverse to the surface
 so we have two relevant extrinsic curvatures. These coincide with the equations of motion for a minimal area surface.
 There are two transverse directions to the surface and thus two equations.
 In appendix B we derive \msec\ directly in $D$ dimensions, without doing the dimensional reduction.

 Note that the non-trace part of the extrinsic curvatures are {\it not } constrained to vanish.
 In fact, already in our simple example of section 2 we have non-vanishing extrinsic curvature if we interpret
 the scalar field as coming from a component of a higher dimensional metric as in \metrcto .\foot{ More
 explicitly, in the notation of that section, if the field at the origin goes as $\phi = r e^{i\tau}$ this leads
 to $\phi_1 = r \cos \tau =x^1$ and there is an extrinsic curvature component $K^1_{y^1y^1} =- K^{1}_{y^3y^3}$ using the
 coordinates \metrcto\ and \btz\ for the 3-d part. }

\subsec{ Computation of the entropy using the
cone method }

Once we have established the form of the solution, we can compute the entropy
using the cone method as explained in section 3.
The arguments are similar, but one has to check that the mild singularities we discussed above cause no problems.

Let us discuss this first for the case of $AdS$ plus a scalar field discussed in section 2.
There, the singularity is only present in the scalar field which behaves as $\phi \sim r^{ \epsilon}  r e^{ i \tau} $.
With this mild singularity, if we integrate by parts in order to use the equation of motion for $\phi$, it is clear
that we do not run into any problem at $r=0$. The most dangerous term seems to come from the variation of
$\delta  \int dr r ( \partial_r \phi )^2  = 2 \int dr r \partial_r \phi  \partial_r \delta \phi \to r \partial_r \phi
 \delta \phi |_0 $. However, $\delta \phi$ would also vanish at the origin if we are considering the variations
  that come from varying $n$ in the solution. In other words, when we compare the correct configuration with $n-1>0$ and
  a regularized cone, we can
  consider a regularized cone where $\phi$ has the same type of singularity at the origin.
  This shows that the first parenthesis in \etnr\ vanishes.

  The second parenthesis only gives us something interesting if we consider terms that have two derivatives acting on the
  metric, otherwise their contribution is going to be small as we remove the regulator. Thus, only the metric in the two
  directions transverse to the minimal surface are relevant. And in those dimensions the computation reduces to the usual
  one, with the contribution coming only from the curvature term in the action.

In conclusion, evaluating the differences in \etnr\ we find that the answer is equal to the area, as we wanted to
prove to argue \RTconj .

The discussion so far was completely local in the directions transverse to the ``horizon''. Here by horizon we
mean the point in the two transverse directions where the circle is shrinking to zero size.
In some cases this ``horizon''  can have  multiple disconnected regions. Then,
 we should sum over the areas of each of the horizons. Even when we have multiple horizons,
the period of the $\tau$ circle is the same in the whole solution.

\subsec{A comments on other $U(1)$ symmetries}

Throughout this discussion we have focused on the  particular geometric circle that we used to define the
 density matrix and the replica trick. We considered cases where we have no  translation symmetry along
 the circle.  However,
we can have other $U(1)$ symmetries. As a simple example, we can have a  $U(1)$ gauge field in the bulk.
Then as in the ordinary case, \GibbonsUE , the integral of the gauge field along the $\tau$ circle should
vanish at the origin $ \left. \int_0^{ 2 \pi n}  d\tau A_\tau \right|_{r \to 0 } = 0 $  (up to global gauge transformations),
where the $\tau$ circle shrinks.
This should hold for all $n$, both integer and non-integer. At the boundary we can fix the
holonomy of $A$ along  the $\tau$ circle as we please. In order to compute the entropy of the density
matrix with a chemical potential we should  fix the integral
$ \hat \mu \equiv \int_0^{ 2 \pi } A_\tau $ at the large $r$ boundary. If we keep everything else fixed at the boundary
but we vary $\hat \mu$, this has the interpretation of changing the density matrix $\rho \to  e^{ i \hat \mu Q} \rho $
where $Q$ is the charge associated to the $U(1)$  symmetry.
We can compute the entropy of this density matrix by treating this boundary condition
 as we treated all other boundary conditions. Namely,
  $A_\tau(\tau)$ is kept fixed. Therefore its circulation over
 the $\tau$ circle of length $2\pi n$ is  $n\hat \mu$.
 Of course, if the holonomy in the $\tau$ circle is different at the origin ($r=0$) than at the boundary, then
 we will have a non-zero field strength in the bulk. The computation of the entropy is identical to what we
 discussed in general.

 This other $U(1)$ symmetry can also be an ordinary geometric isometry, and its treatment is similar.

\newsec{ Connection to the Ryu-Takayanagi formula }

We  presented
the computation of the entropy of the gravitational density matrix  in a form that is very general.
The objective was to emphasize that \RTconj\ is really a statement about an analytic continuation
of classical solutions.
In this section we explain why the  conjecture \RTconj\ for the
  entropy is related to the Ryu-Takayanagi  formula for entanglement entropy.

The Ryu-Takayanagi formula is a conjecture in the  AdS/CFT context \RyuBV .
 In the quantum field theory
one is interested in computing the entanglement entropy of a spatial region $A$ on the boundary
of the field theory. This spatial region has a boundary $\partial A$. The conjecture is that
this entanglement entropy is given by a the area (in Planck units) of a codimension two minimal
surface in the bulk whose boundary ends on $\partial A $.
\ifig\rt{ The Ryu-Takayanagi conjecture. The entanglement entropy of a region $A$ in a conformal field
theory is given by the area of a minimal surface in the bulk of $AdS$ that ends on $\partial A$
(the boundary of region
 $A$) at the boundary of $AdS$. } {\epsfxsize1.5in\epsfbox{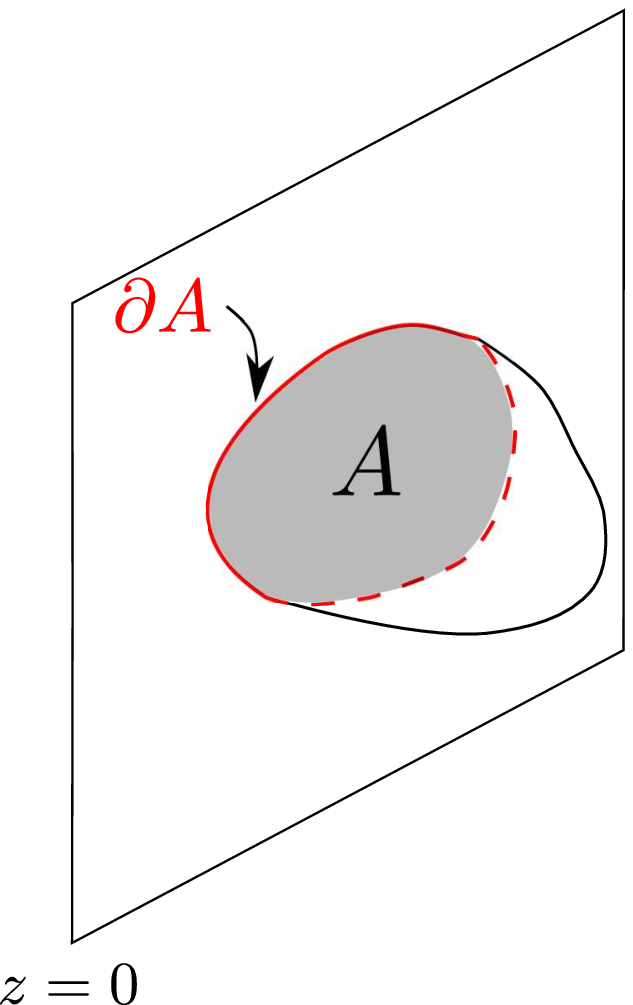}}

In principle, we can compute the entanglement entropy of the region $A$ by using the replica trick  \refs{\CallanPY,\CalabreseEU}.
This is a general method for computing entanglement entropy in quantum field theories.
The idea is to take $n$ copies of the field theory and match them together so that by moving in a circle around
$\partial A $ we go from one copy to the next. Going $n$ times around this circle we come back to the original copy.
Thus at $\partial A$ there is a conical defect with a $2\pi n$ opening angle.
This appears to be a singular metric. However, one can choose a conformal factor that diverges at $\partial A$ in
such a way that the size of the circle around $\partial A$ is finite.

\ifig\rtb{
   Geometries that we are considering to compute the entanglement entropy in the field theory. a)
Semiplane. Region $A$ is half of the plane and its boundary, $\partial A$, is at
 $x_0=x_1=0$.  b) $x_0, x_1$ view of the semiplane and the coordinate $\tau$. }  {\epsfxsize4.5in\epsfbox{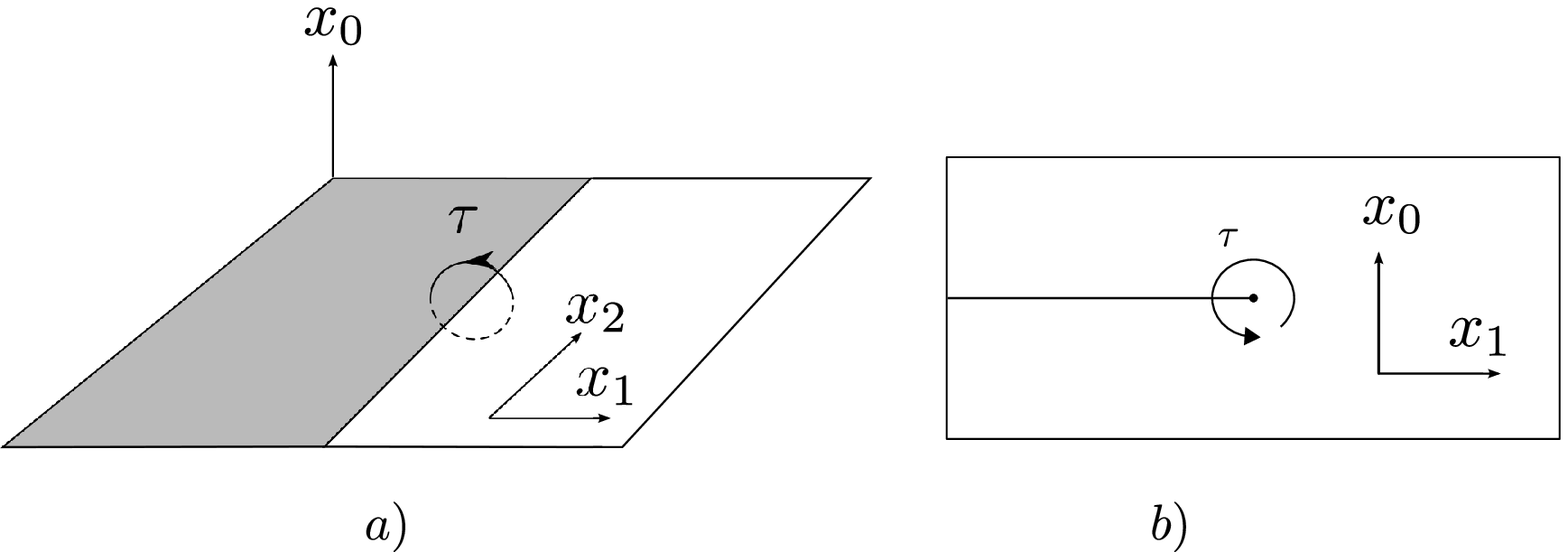}}
\ifig\rtc{
 Other geometries in three dimensions. a)
Disk configuration.  b) Slightly deformed disk. The $\tau$ coordinate goes around  $\partial A $.  } {\epsfxsize4.5in\epsfbox{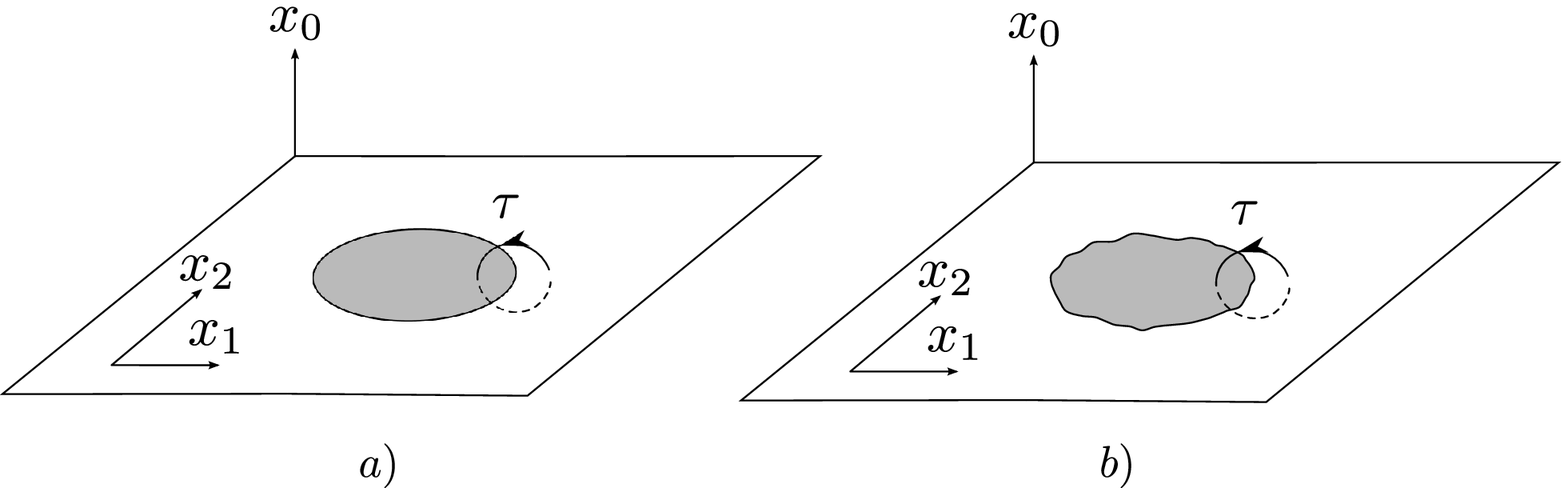}}

This is most easily understood for simple regions \refs{\CasiniKV,\HungNU}.
 Imagine we have a  conformal field theory
 $R^{1,d-1}$ at the boundary. Then we can
choose a region $A$ defined by  $x_1>0$, see figure \rtb .
 The boundary of the region is the surface $x_0=0$, $x_1=0$. Going to Euclidean space, $R^{d}$,
we can combine the directions $x_0$ and $x_1$ into two directions labeled in polar coordinates by $r$ and $\tau$.
The metric is
\eqn\metrc{
ds^2 =   r^2 d\tau^2 + dr^2 +   d \vec x^2   ~~~~ \to ~~~   d\tau^2 + { dr^2 +  d \vec x^2 \over r^2 }
}
where $\vec x$ are the rest of the spatial coordinates. In the right hand side of \metrc\ we have multiplied
by an overall conformal factor $1/r^2$ to put the metric in the form of $S^1 \times H_{d-1}$. We can now easily
perform the replica trick, it corresponds to changing the length  of $S^1$ from $2\pi$  to $ 2 \pi n$.
Clearly this metric, $S^1_n \times H_{d-1}$ is a perfectly legal metric and we can consider its gravity dual. It is
a certain black brane. In this case, we have a $U(1)$ isometry in the rescaled coordinates and then the entropy
computed using the replica trick or using the ordinary Gibbons-Hawking formula is exactly the same. Note that at the
$AdS$ boundary the circle $S^1$ has a nonzero size everywhere.
In the interior of $AdS$ it shrinks to zero at a ``horizon''.
Notice, in particular, that the half space region we discussed above can be conformally mapped to
a spherical region $ \sum x_i^2 \leq 1 $. In this case, the circle $S^1$ appearing in \metrc\ corresponds
to a coordinates that goes around $\partial A $ as in figure \rtc .

Now, this was a very simple region. If we consider more complicated regions,
 then it is not possible to choose a
system of coordinates and a conformal rescaling such that the metric is
 independent of the angular direction $\tau$.
In all cases we will have an angular direction, $\tau$,
 since it is the direction we used to perform the replica trick construction. The choice of this coordinate
is completely arbitrary, as long as it goes around the boundary of region $A$.
As we go near $\partial A$ we have a problem which locally looks like \metrc , and we can
choose a conformal factor which makes the metric non-singular as in \metrc\ for all the replicas.
The difference with \metrc\ is that, as we increase $r$, we will have extra terms in the metric that can have
some $\tau$ dependence. This dependence  always involves powers of $e^{ \pm i\tau}$ since this is just the statement
that the $\tau$ direction is parametrizing circles in the original boundary geometry.
 All the statements in this paragraph
involve the boundary geometry, the geometry where the field theory lives. These  replica trick boundary geometries simply
amount to letting the circle have size $\tau \sim \tau + 2 \pi n $, {\it without changing any of the functions that
appear in the boundary geometry}. All such functions are periodic under $\tau \to \tau + 2 \pi $ \foot{
If we want to explicitly  parameterize the metric in this way, we might need to choose different coordinate patches, as usual.  When the coordinate patches are chosen in a $\tau$ dependent fashion,
then the $\tau \to \tau + 2 \pi n$ identification can produces spaces with an $n$ dependent
topology. This
happens, for example, in the case that we have two separate intervals in a two dimensional CFT.
(We thank Xi Dong for a discussion on this.) }.

Thus, the field theory replica trick, can be translated, via the  standard AdS/CFT
dictionary \refs{\GubserBC,\WittenQJ}, to
a problem in gravity which is identical to the problem that we discussed in section 4. Here no conjecture is involved
other than the original AdS/CFT relation. The replica trick then defines the entropy as in \entrog . In order to do
that,  we need to analytically continue in $n$ to $n\sim 1$.

The Ryu-Takayanagi conjecture boils down to a statement in classical geometry. It is the statement we discussed
in section 4. Computing $\log Z(n)$, using smooth geometries,   analytically continuing in $n$, and
computing the entropy defined in \entrog\ gives the area formula in \RTconj .

Notice that in a setup where $A$ is a spatial region contained at $x^0=0$ on the boundary, then there is a time
reflection symmetry $x^0 \to - x^0$, which translates into $\tau \to -\tau$ for the circle in the Euclidean solution.
This implies that we can go to Lorentzian signature, as usual with $x^0 \to i x^0$. This translates into
$\tau \to i t $ . Now the region where the $\tau$ circle is shrinking to zero corresponds to a horizon in the bulk. It is a horizon for an observer sitting at fixed small $r$.

There is a generalization of the Ryu-Takayanagi conjecture for situations
 that are time dependent \HubenyXT .
It again involves an extremal surface ending on $\partial A$, but in the full Lorentzian spacetime.
In those cases
there is no obvious Euclidean continuation to perform the replica trick. This suggests that there should
be a way to think about the problem which does not go through the Euclidean solutions and the replica trick.
We should remark that in some cases we can perform a replica trick in the Euclidean geometry for regions
that depend on the Euclidean time and then one can analytically
continue to the Lorentzian signature solutions. Some examples were discussed in \HartmanQMA .

\subsec{General entanglement interpretation}

In the introduction, we presented the computation of the generalized gravitational entropy as
a property of the density matrix constructed by integrating over a circle in Euclidean time.
It is natural to ask whether there is a general Lorentzian interpretation that involves entanglement.
This is indeed  the case   in the Ryu-Takayanagi discussion of entanglement of a subregion of the boundary.

\ifig\rhoandpsi{ We consider periodic boundary conditions with a reflection symmetry $\tau \to -\tau$.
  In (a) we see that by cutting at $\tau=0$ we get a density matrix $\rho_{ac}$, where
  $a$ and $c$ label the states on the two sides of the cut. In (b) we note that we can cut along the moment of
  time reflection symmetry $\tau=0, \pi$. Then we get a pure state in two separate Hilbert spaces labeled by A and B.
  The bottom half of the picture can be viewed as a state $\psi_{ab}$ and the top part as $\psi^\dagger_{cd} $.
  Tracing out over the $B$ Hilbert space, we recover $\rho_{ac} = \sum_b \psi_{ab} \psi^\dagger_{cb} $. At this
  moment of time reflection symmetry we can also continue to Lorentzian signature.  } {\epsfxsize3.5in\epsfbox{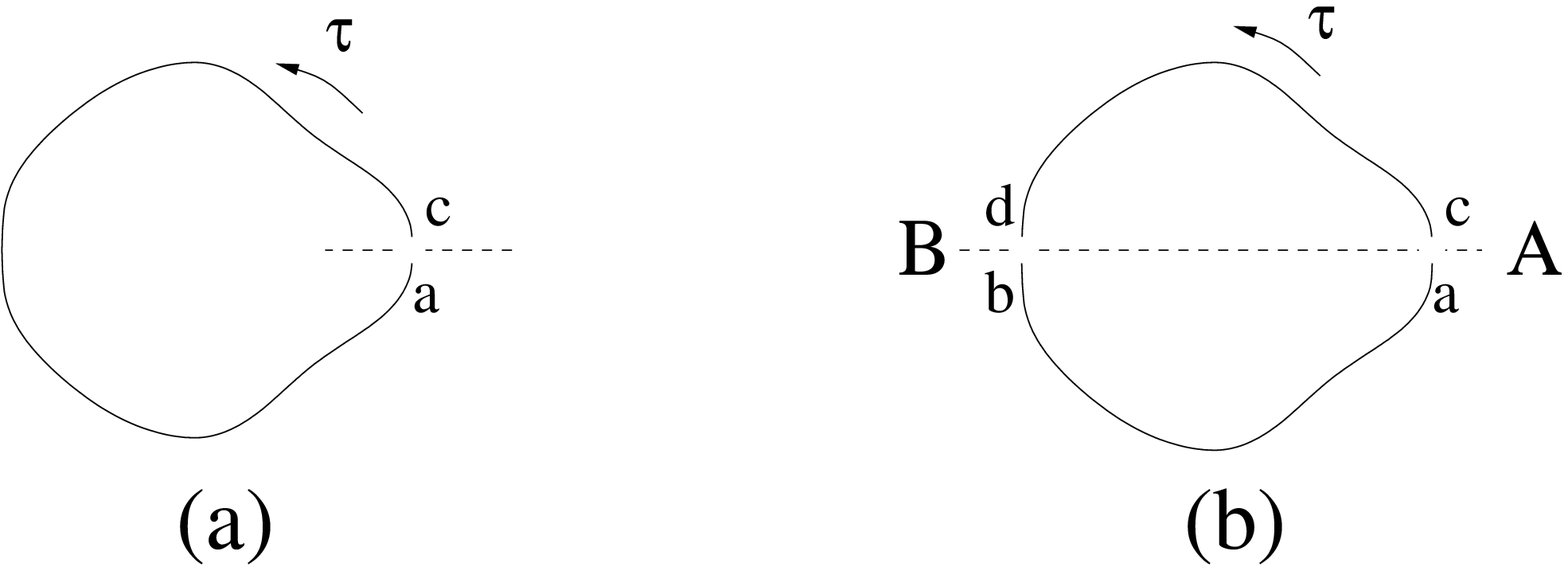}}

Here we would like to point out that in very general situations we can also have an entanglement interpretation. Suppose that the boundary conditions have a moment of time reflection symmetry.
Say that this acts as $\tau \to -\tau$.
Then by cutting the boundary conditions at $\tau =0, \pi$ we can interpret the lower part of the evolution as
specifying a pure state $|\Psi\rangle$ in the product of two theories, which we call $A$ and $B$. See \rhoandpsi .
 Similarly, the upper part can be
viewed as specifying the state $\langle \Psi | $.  The density matrix can then arise by tracing over
one or the other subsystem. And the entropy can be interpreted as entanglement entropy for system $A$ with  $B$.
This is the same as in the eternal black hole discussion \refs{\IsraelUR,\MaldacenaKR}.

\ifig\Lorentzian{ Here we consider a situation with asymptotically AdS boundary conditions. The boundary conditions
contain a small time dependent deformation which vanishes at infinity. So in the far future we settle down into a
stationary black hole on both sides. The entropy of these black holes is bigger than the entropy of the initial entanglement since, the time dependent boundary conditions have sent in energy and have increased the entropy
of the system. In other words, there was a non-zero flux of energy through the horizon which increased its area.
The dotted lines indicate the matter falling through the horizon.
  } {\epsfxsize2.5in\epsfbox{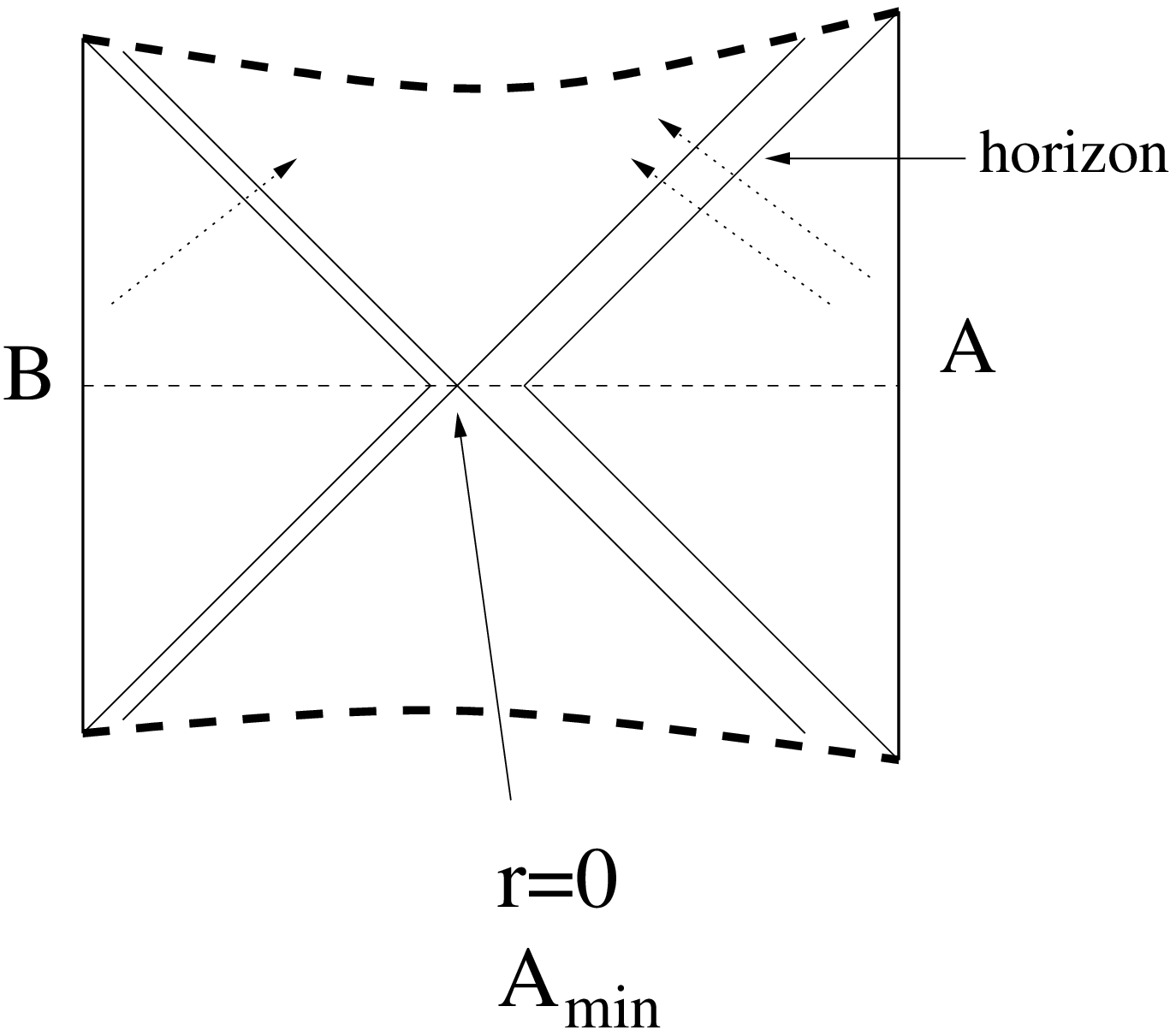}}

The bulk solution is also expected to have a time reflection symmetry in this case.  Under $\tau \to i t$ we
get a Lorentzian solution. The vicinity of $r=0$ looks locally like Rindler space.
This procedure generically produces a time
dependent solution and we might get singularities or horizons in the boundary conditions. We can
 consider a situation where the Lorentzian time evolution can be performed out to infinite time
without ever connecting again the two boundary regions or encountering singularities on the boundary.
An example is the following.  We start from a Euclidean black hole but with a
small perturbation of the boundary conditions which is smooth in Euclidean time and goes to zero at large Lorentzian
time. More concretely,
 we can consider the model of section 2 and set the boundary conditions $\phi_B ={  \eta  (1+ \cos \tau)
\over 2 + \cos 2 \tau } $. When we go to Lorentzian time this becomes $\phi_B ={\eta (1 \pm \cosh \tau)  \over 2 + \cosh 2 t } $ where the $\pm$ corresponds to the A and B sides respectively. Note that these go to
zero at large times. We expect that solution should be qualitatively like \Lorentzian . A very
explicit solution with these characteristics was studied in  \refs{\BakJM,\BakQW} \foot{ The solutions in \refs{\BakJM,\BakQW} are based on Janus solutions. Their
   boundary in Euclidean space has the form $S^1 \times \Sigma$ where $\Sigma$ is
  a quotient of hyperbolic space. The $S^1$ is divided in two equal parts and the dilaton has
  a different value on each part. The Lorentzian continuation is obtained by continuing across the
  moment with a time reflection symmetry. The two boundaries different values for the dilaton. These values
  are constant in time. The bulk smoothly interpolates between the two.  }.

In cases that arise from  entanglement of subregions via AdS/CFT,
 the fact that the causal horizon is closer to the boundary
than the minimal surface that computes the entanglement entropy was noted in \HubenyXT (see also
 \refs{\FreivogelQH,\HubenyWA,\CzechBH,\BoussoSJ}).

\newsec{Conclusions and  discussion }

In this article we have noted that we can generalize the concept of Euclidean gravitational entropy
to more general situations than the ones associated to thermal equilibrium.
In particular, we have considered euclidean solutions that contain a circle $\tau \to \tau + 2 \pi $.
We have introduced a boundary, setting boundary conditions which  are $\tau$
dependent but  periodic under $\tau \to \tau + 2 \pi $. Thinking of gravity as a holographic
theory, we view these boundary conditions as defining the system. Euclidean evolution on the circle produces an
un-normalized density matrix. The Euclidean solution gives us the trace of this density matrix.
By performing the gravity version of the replica trick we have defined traces of $n^{th}$ powers of the
density matrix. These are geometries with exactly the same boundary condition as functions of $\tau$,
but where the $\tau$ variable is taken
to have period $\tau \to \tau + 2 \pi n$. For integer $n$ the bulk geometries are smooth and free of any conical
defects. These geometries are computing the trace of the $n^{th}$ power of the density matrix.
By analytically continuing in $n$ and taking a derivative near $n=1$ we can compute a quantity that is
interpreted as the Von-Neumann entropy of the underlying density matrix.
Note that all computations are classical.
 The density matrix we are talking about is a hypothetical density matrix
in some underlying theory of quantum gravity. In AdS/CFT situations we can give an precise definition for this
density matrix.

A version of the Ryu-Takayanagi conjecture is that this generalized gravitational entropy, computed in this fashion, is
given by the area of a minimal area surface in the original geometry (the solution with $n=1$).

We have given some arguments for the correctness of the Ryu-Takayanagi conjecture. The arguments involved
the assumption that we can analytically continue the geometries away from integer values of $n$.
We further made the assumption that these analytically continued geometries, for small $\epsilon \equiv  n-1$,
are smooth in the two directions transverse to the minimal area surface but can have mild singularities which
are not important for evaluating the action. We do not view these metrics as physically meaningful, we view
them just as a tool for deriving the Ryu-Takayanagi formula.
Our assumptions were motivated by considering a simple example, described in section 2.  But we  have no
further justification other than the fact that they hold in this example and seem reasonable assumptions.
We have derived the minimal area condition by  demanding the existence of a small deviation away from
the $n=1$ solution  that is consistent with our assumptions on the type of singularities that are allowed.
One simple way to state the type of allowed singularities is to do a dimensional reduction of the whole configuration
to the two dimensions transverse to the minimal surface. Then we have a two dimensional metric, a dilaton field that
multiplies the two dimensional curvature in the action and a set of other fields. Then the metric should be smooth and
the gradient of the dilaton at the minimal surface should be zero, which is the minimal area condition.
All other fields can have mild singularities of the form $ \phi \sim  z |z|^{ 2 \epsilon}$ at the origin.
  When $n$ is not an integer we evaluate the gravitational action by integrating $\tau$ between $[0,2\pi]$ and then multiplying by $n$.
We have also argued that this method gives rise to the area formula for the entropy, essentially for the same reasons
as for the case with the $U(1)$ symmetry. One way to understand this is that all non $U(1)$invariant fields are going
to zero at the origin.  Then the methods described in section 3 give the usual formula.

An alternative way to view the solutions is to imagine that we keep the original period of the circle,
$\tau \sim \tau + 2 \pi$ but we introduce a cosmic string  (or cosmic $D-3$ brane) with a
$2 \pi/n$ opening angle. In addition,  we multiply the resulting action by a factor of $n$. For $n$ close to
one we have a very light cosmic string that deforms the geometry very slightly.  We  can then view
the entropy formula as arising from the Nambu action for this cosmic string.  Also the
minimal area condition comes from minimizing this Nambu action. The long and detailed discussion
that we presented tried to justify these statements in detail.

One interesting open question is whether one can generalize the derivation to the time dependent
case considered in  \HubenyXT , where, generically,
 there is no obvious Euclidean continuation.

Another interesting direction is to generalize the discussion to gravity with higher derivatives. The most
naive conjecture is that
the entropy is given by the Wald formula. However, this conjecture was argued to be wrong in \HungXB , where a modified conjecture was made for the case of Lovelock gravity.
A more informed conjecture is  to say that we get the Wald-Iyer formula proposed in section 7 of
\IyerYS . In fact, this reduces to the proposal in \HungXB\ for  Lovelock gravity. It would be interesting
to see whether this is correct and what the equations for the surface  are.

\noindent{\bf Acknowledgments}

\noindent
We  thank  T. Faulkner, T. Hartman and I. Klebanov for discussions.

 \noindent
This work was supported in part by U.S.~Department of Energy grant DE-FG02-90ER40542.  A.L. acknowledges support from ``Fundacion La Caixa".

\appendix{A}{Example of a scalar field in $AdS_3$}

In this appendix, we consider a massive scalar field in $AdS_3$ and show explicitly that
 the  entropy that we compute using the replica trick
  is equal to the modification of the area due to the presence of a non-zero  scalar field background.

\subsec{Massive scalar field }

For a massive scalar field we have equations which are very similar to the ones in the text. We consider a
complex scalar field of mass $m$. Setting the radius of $AdS_3$ to one we need to impose the boundary condition
\eqn\bodco{
\phi|_{r_c}  = \eta e^{ i \tau } r_c^{ \Delta - 2 } ~,~~~~~~~~~~~~\Delta = 1 + \sqrt{ m^2 + 1 }
}
where $\Delta$ is the scaling dimension of the corresponding operator and $r_c$ is a large value of $r$ which represents
the cutoff surface.
The relevant solution of the wave equation on the metric \intb\  is
\eqn\solwva{
\phi = \eta e^{ i \tau } \frac{f( n r )}{f(n r_c)} r_c^{\Delta -2}  ~,~~~~~~~~~~~~f ( r ) = r^{n}   \, _2 F_1\left(\frac{n}{2}-{ \Delta
\over 2}  +1,\frac{n}{2}+{ \Delta \over 2 }  ;n+1;-r^2\right)
}
We then evaluate
\eqn\sacc{\eqalign{
\log Z(n)= & -  \int d^3x \sqrt{g} [ | \nabla \phi |^2+m^2 |\phi|^2 ]  =
 - (2 \pi n) L_x  \left.   \phi^{*}_{r_c} r_c^3 \partial_r \phi\right|_{r_c} =
 \cr
 = &  ( 2 \pi  L_x ) |\eta|^2  \left[ B(n,\Delta)  + {\rm linear~in~}n \right]
 }}
 where the terms linear in $n$ also include all divergent terms. It is important that these counterterms do not give
 rise to any non-trivial $n$ dependence. This is due to the fact that we keep the $\tau$-dependence of the boundary conditions
 fixed as we vary $n$.
We also defined

\eqn\bn{
B(n,\Delta)=-\frac{2 n^{3-2\Delta} \Gamma (2-\Delta ) \Gamma \left(\frac{n+\Delta }{2}\right)^2}{\Gamma (\Delta -1) \Gamma({n-\Delta \over 2}+1)^2} }
We can then compute the entropy to order $\eta^2$ from \entrog , which gives
\eqn\entr{ \eqalign{
S|_{\eta^2} = & -   n\partial_n [ \log Z(n) - n \log Z(1) ]|_{n=1} =
 \cr
 = & - \eta^2  \left \{ \frac{4 \pi  \left[ 2 (\Delta -2) \Delta +   (1 -  \Delta )\pi  \tan (\pi  \Delta/2 ) \right] \Gamma (2-  \Delta ) \Gamma \left({ \Delta +1 \over 2 }  \right)^2}{\Gamma \left(\frac{3-\Delta}{2}  \right)^2 \Gamma (  \Delta )}
\right \}
}}

\subsec{Change in the metric from Einstein's equations}

Now we will study the backreaction of the scalar in the metric. The action is
\eqn\lagsc{
-S=\int_{AdS_3} \left [R-2 \Lambda-| \nabla \phi |^2+m^2 |\phi|^2 \right ]
}
with $\Lambda=-1$. The equations of motion are
\eqn\eeq{
R_{\mu \nu}-\frac{g_{\mu \nu}}{2} (R-2)= T_{(\mu \nu)}
}
where $T_{\mu \nu}=\partial_{\mu} \phi^{*} \partial_{\nu} \phi-\frac{g_{\mu \nu}}{2} (| \nabla \phi |^2+m^2 |\phi|^2)$.
The ansatz for the metric is
\eqn\pertmetric{
ds^2= \frac{1}{r^2+  g(r)+1} dr^2+\left (r^2 +1\right )(1 + v(r) ) d x^2+ r^2 d t^2
}
where $g(r),v(r)$ will be $O(\eta^2)$. If we expand Einstein equations to first order we obtain three
 equations for the diagonal components.
There are only two independent equations since the last one will give us the scalar wave equation when the first two are satisfied
\eqn\Eineqs{\eqalign{
  g'(r)=T_{xx}\frac{2 r }{\left(r^2+1\right)} \cr
  v'(r)=2 r T_{rr}-\frac{2 r g(r)}{\left(r^2+1\right)^2} \cr
}}




Since we consider a configuration with
  $\partial_x \phi=0$,  we can relate the components of the stress energy tensor: $T_{rr}=(\partial_r \phi)^2 +  (1+r^2)^{-2}  T_{xx}$. We then find
\eqn\veq{
v'(r)=2 r (\partial_r \phi)^2+{\partial \over \partial r} \left( {g(r) \over r^2+1} \right)
}
And
\eqn\intg{
v(0) = - 2  \int_0^\infty d r r | \partial_r \phi |^2 ~~~ \to   ~~~~~~~S|_{\eta^2}  = {4\pi \delta A } =
- \eta^2 ( 4 \pi L_x)   \int_0^\infty d r r   |\partial_r \phi(r)|^2
}
where we use that the second term in \veq\ is a total derivative and that $g(0) =0$ due to the regularity condition
for the metric at the origin.  In addition $g/r^2 \to 0$ at infinity.
In our units ($16 \pi G_N=1$),  the black hole formula is $S=4 \pi A=4 \pi A_0(1+{ v(0)  \over 2})=4 \pi ( A_0+\delta A)$.
Substituting the solution for $\phi(r)$ for $n=1$ \solwva , and integrating, we get the same as in \entr .
We checked this only numerically, but below we will show it without performing the explicit calculation.

\subsec{The two quantities are the same}

In the above computation we actually did not need to solve all the equations to the end in order to
show that the two results are the same.

We will rearrange the entropy formula for the scalar so that  we get an expression that is simpler to compare with the area contribution. The lagrangian $  {\cal{L}}(g_{\mu \nu},\phi,\nabla_{\mu} \phi) $ is a function of
$\tau$. When we evaluate the gravitational action, we integrate over all coordinates except $\tau$.
Then we first  integrate over $\tau$ from zero to $2 \pi$ and then
multiply by $n$. We can do this both for integer or non-integer $n$.
We denote the $\tau$ integral as $ [ \log Z(n) ]_{2 \pi} $. Then we have
\eqn\logzn{
\log Z(n) = n [ \log Z(n) ]_{2 \pi}
}
Then the entropy formula \entrog\ simplifies and we get
\eqn\tauavg{S =- \left.  n \partial_n \left\{   n[ \log Z(n) ]_{2 \pi} - n \log Z(1) \right\}\right|_{n=1}=-
\left.
\partial_n [ \log Z(n) ]_{2 \pi} \right|_{n=1}
}
And the later expression can be straightforwardly evaluated, using\foot{We define it like this because the action is $\log Z(n)=\int_{AdS_3}  ({\cal L}_{\rm Grav}-{\cal L}_{\rm matter})$ so the field equations read $G_{\mu \nu}=T_{\mu \nu}$. }
$\sqrt{g} T_{\mu \nu}={\partial \sqrt{g} {\cal{L}} \over \partial g^{\mu \nu}}$ ,
\eqn\EEscala{\eqalign{
-\partial_n [ \log Z^{\rm matter}(n) ]_{2 \pi}
 = &  \int_0^{2 \pi}  d\tau  \int dx dr \sqrt{g} \left (T_{\mu \nu}\frac{\partial g^{\mu \nu}}{\partial n}+ \frac{\partial \cal{L}}{\partial \phi} \partial_n \phi+ \frac{\partial \cal{L}}{\partial (\partial_{\mu}\phi) } \partial_n \partial^{\mu} \phi \right )=
\cr =& \int_0^{2 \pi }  d\tau \int dxdr  \sqrt{g} T_{\mu \nu}\frac{\partial g^{\mu \nu}}{\partial n}
}}
 In the last line we used the equations of motion (of course $\frac{\delta S}{\delta \phi}=0$). One can check that the expression with the stress energy tensor gives us $\sqrt{g} T_{\mu \nu}\frac{\partial g^{\mu \nu}}{\partial n} |_{n=1}=-2 \eta^2 r f'^2$, so
\eqn\EEscalb{
S-S_0=-\eta^2 4 \pi L_x  \int dr r f'^2
}
In writing \EEscala\ we have only included the action of the scalar field in the computation.

We can now show that we get the area, without using explicit expressions. This can be done
as follows. First note that in the second line of \EEscala\ we can use Einstein's equation to write
 $T_{\mu\nu}$ in terms of the Einstein tensor, which is related to the variation of the gravitational
action. We end up with an expression of the form
\eqn\eeng{ \left.
 \int d \tau dx dr \sqrt{g} G_{\mu\nu} {
 \partial g^{\mu\nu } \over \partial n } \right|_{n=1}
}
 This is closely related to the derivative of the gravitational
part of the action.
As we explained above
we know that the gravitational part of the action has no term of order $\eta^2$.
Thus we know that the $\partial_n$ derivative of the gravitational part vanishes at order $\eta^2$.
This derivative is the same as  \eeng\ up to a total derivative term
\eqn\sgrg{
 \left. \partial_n [ \log Z^{\rm Grav}(n) ]_{2\pi} \right|_{\eta^2}   =0 = 2 \pi \left[
 \int dx dr \sqrt{g} G_{\mu\nu} {
 \partial g^{\mu\nu } \over \partial n }  - \left. \int dx  \sqrt{g} \,   \nabla_\mu \partial_n g^{\mu r} \right|_{r=0}
\right]_{\eta^2}
}
The last term gives the area of the horizon, or more precisely the area of the horizon at order $\eta^2$.

These are the same manipulations that one can do in general, but we have done all steps explicitly above
to check that everything  indeed works in situations with no U(1) symmetry.

\subsec{ Real scalar}

We now consider the case of a
 real scalar field $\phi=f(r) \cos  \tau$, $f(r)$ is the same as before but now
 the stress energy tensor  no longer has the $U(1)$ symmetry
\eqn\tmunureal{
T_{\mu \nu}=T^{0}_{\mu \nu}+T^1_{\mu \nu} e^{i 2  \tau}+T^{-1}_{\mu \nu} e^{-i 2  \tau}
}
And $T_1^*=T_{-1}$. The metric has the same fourier decomposition, so $v$ and $g$ in \pertmetric\
also have three fourier components.
The entropy coming from the change in the area is
\eqn\EEreal{S|_{\eta^2}=\int_0^{2 \pi} d \tau v(0)=2 \pi v^0(0)}
where $v^0$ is the constant component of $v$. It is easy to check that $v^0(0)=-\int dr r f'^2$.

The scalar action contributes as follows
\eqn\EEscalareal{S|_{\eta^2}=\int d \tau dr(-2 r f'^2 \cos^2 (k \tau))=-2\pi \int dr r f'^2  }
So we find agreement once more, and the result is precisely half of the complex scalar.

\appendix{B}{Derivation of minimal area condition for the general case from a explicit calculation}

In this appendix, we   obtain the minimal area condition of section 4 without using dimensional reduction. As in section 4, we   derive this condition from requiring that the analytically continued solution satisfies the linearized equations of motion near $r=0$.

The metric of the $n=1$ solution, which satisfies (locally) the equations of motion is
\eqn\expmeb{\eqalign{
ds^2  = & dx_1^2+dx_2^2 +  g_{ij}  ( dy^i + b^i_\alpha dx^\alpha ) (dy^j + b^j_\alpha dx^\alpha )  +o(r^2)~,
\cr
 g_{ij} = &
 h_{ij} +  x_1 K^1_{ij} + x_2 K^2_{ij} ~,~~~~~~~~~ b^i_\alpha \sim o(r)
 }}
Here,   $y_i$ are the directions along the surface. Now, we do the replica trick, that is, we change the periodicity of the $\tau$ circle from $2\pi$ to $2\pi n$ and analytically continue $n$ to $1+\epsilon$. In this way, the metric will be modified to linear order in $\epsilon$
\eqn\metfniten{
ds^2  = e^{2 \rho} (dr^2+r^2 d\tau^2) +  g_{ij} ( dy^i + b^i_\alpha dx^\alpha ) (dy^j + b^j_\alpha dx^\alpha ) + \delta g
}
Where we decomposed the perturbation in a part that makes the metric smooth
$\rho = \delta \rho = - \epsilon \log r  $  and a perturbation $\delta g$  that has components $\delta g_{a b}$ valued in all directions. For simplicity we work with $z, \bar{z}$ coordinates: $x_1={ z+\bar{z} \over 2},x_2={z-\bar{z} \over 2 i}$.
As a gauge condition, we  set $\delta g_{z z}=\delta g_{\bar{z} \bar{z}}=0$.
 We also set $\delta g_{z \bar z} =0$, since this variation is included in $\rho$.  We require the  perturbation, $\delta g_{a b}$ to be periodic: $\delta g_{a b}(\tau) \sim \delta g_{a b}(\tau+2\pi)$.

We want to compute the linearized equation of motion $\delta G_{zz} = \delta T_{zz}$.
In particular, we want to focus on the terms that can be divergent, going like $1/r$ near the origin.
 We find

\eqn\flucEin{\eqalign{ \delta R_{z z}= &
\frac{- \epsilon}{z} K_z  + \frac{1}{2} (2 {\delta g^{p}}_{ z;z p}-\delta g_{; z z}-\nabla^2 \delta g_{z z})  +
 (\text{regular ~ as ~~} r \to 0)
\cr
= &  \frac{- \epsilon}{z} K_z   -\frac{1}{2} \partial^2_{z} \delta \gamma + \cdots
}  }
where $\delta \gamma \equiv g^{i j} \delta g_{i j}$ and $K_z={K^1-i K^2 \over 2}$ .
In \flucEin\ we neglected the terms that have $y_i$ derivatives because we expect them to be regular, only terms
 with two $x^\alpha$ derivatives can contribute to this order.

 Now, since the stress energy tensor is not expected to be singular,  the equations of motion
 imply that the     two potentially divergent terms should cancel
\eqn\EOMR{\eqalign{
\frac{1}{2} \partial^2_{z} \delta \gamma= \frac{- \epsilon}{z} K_z \cr
\frac{1}{2} \partial^2_{\bar{z}} \delta \gamma= \frac{- \epsilon}{\bar{z}} K_{\bar{z}} }}

These are the same equations as before \condc , which are only satisfied for  a periodic function,
$\delta \gamma(\tau) \sim \delta \gamma (\tau+2\pi)$, if $K_z=K_{\bar{z}}=0$. Note that although the equations of motion are well behaved for $K_z=0$, the Riemann tensor diverges, as we discussed in section 2. This discussion is similar
to the analysis in \UnruhHY\ for the motion of a cosmic string.

\appendix{C}{ Computation of the entropy for a disk }

Here we consider a very simple example of gravitational entropy. We go through it to explain how one
can put boundary conditions at fixed distance.

Consider the metric  $ds^2 = dr^2 + r^2 d\tau^2 $.
In addition, we can have other dimensions, but let us assume we can ignore them.
In this case,  we can say that we pick an $r=r_c$ and we set up the boundary conditions there. We demand
that the metric in the angular direction is
\eqn\bdcd{
ds^2_{bdy} = r_c^2 d\tau^2
}
at the boundary $r=r_c$. We now consider the situation with $\tau \sim \tau + 2 \pi n$.
We should consider now metrics with the same boundary condition \bdcd , but compatible with the new period.
These metrics are
\eqn\mecgb{
ds^2 = n^2 dr^2 + r^2 d\tau^2
}
We can evaluate the gravitational action for these spaces and obtain
\eqn\grava{
\log Z(n)  = { 1 \over  16 \pi G_N} \left[ \int \sqrt{g} R  + 2\int_{bdy}  K \right]  = { A  \over 4 G_N}
}
which is independent of $n$. Here $A$ is the area of the transverse directions which were not explicitly mentioned above.
Using the usual formula, we get the expected area formula for the entropy.

We have included this trivial computation to explicitly show how gravity regularizes the divergent contribution that
one normally gets in field theory. In fact, there is no divergence because there was no conical space in this
computation!. Of course, this begs the question of whether the finite part of the
one loop correction computed by performing a one
loop computation around the above geometries is indeed the same as the finite part of the
 one loop corrections computed using the
conical spaces that appear in the field theory discussion of the replica trick.

\listrefs

\bye